\documentclass[lettersize,journal]{IEEEtran}
\usepackage{graphicx}
\usepackage{caption}
\usepackage{comment}
\usepackage{textcomp}
\usepackage{subcaption}
\usepackage{float}
\usepackage{placeins}
\usepackage{multirow}
\usepackage[table]{xcolor}
\usepackage{graphicx}
\usepackage{array}
\usepackage[table]{xcolor}
\usepackage{subcaption}
\usepackage{afterpage}
\usepackage{amsmath} 
\usepackage{booktabs} 
\usepackage{adjustbox}
\usepackage[numbers,sort&compress]{natbib}
\usepackage{enumitem} 
\usepackage{hyperref}
\usepackage{orcidlink}
\usepackage{tabularx}

\begin{document}
\title{IoT-based Noise Monitoring\\
using Mobile Nodes for Smart Cities}
\author{
\IEEEauthorblockN{Bhima Sankar Manthina\textsuperscript{1}\orcidlink{0009-0005-9412-7890}, Shreyash Gujar\textsuperscript{1}\orcidlink{0009-0004-8047-3576}, Sachin Chaudhari\textsuperscript{1}\orcidlink{0000-0003-1923-0925}, Kavita Vemuri\textsuperscript{1}, Shivam Chhirolya\textsuperscript{2}\orcidlink{0009-0001-1108-7928}}
\IEEEauthorblockA{\textit{\textsuperscript{1}International Institute of Information Technology-Hyderabad (IIIT-H), India,}
\IEEEauthorblockA{\textit{\textsuperscript{2}Prezent.AI, India} \\
\{bhima.sankar, shreyash.gujar\}@research.iiit.ac.in, \{sachin, kvemuri\}@iiit.ac.in, shivam.chhirolya@prezent.ai}
}}

\maketitle

\begin{abstract}
Urban noise pollution poses a significant threat to public health, yet existing monitoring infrastructures offer limited spatial coverage and adaptability. This paper presents a scalable, low-cost, IoT-based, real-time environmental noise monitoring solution using mobile nodes (\emph{sensor nodes on a moving vehicle}). The system utilizes a low-cost sound sensor integrated with GPS-enabled modules to collect geotagged noise data at one-second intervals. The sound nodes are calibrated against a reference sound level meter in a laboratory setting to ensure accuracy using various machine learning (ML) algorithms such as Simple Linear Regression (SLR), Multiple Linear Regression (MLR), Polynomial Regression (PR), Segmented Regression (SR), Support Vector Regression (SVR), Decision Tree (DT), and Random Forest Regression (RFR). While laboratory calibration demonstrates high accuracy, it is shown that the performance of the nodes degrades during data collection in a moving vehicle. To address this, it is demonstrated that the calibration must be performed on the IoT-based node based on the data collected in a moving environment along with the reference device. Among the employed ML models, RFR achieved the best performance with an \( \mathrm{R}^2 \) of 0.937 and RMSE of 1.09 for mobile calibration. The system was deployed in Hyderabad, India, through three measurement campaigns across 27 days, capturing 436,420 data points. Results highlight temporal and spatial noise variations across weekdays, weekends, and during Diwali. Incorporating vehicular velocity into the calibration significantly improves accuracy. The proposed system demonstrates the potential for widespread deployment of IoT-based noise sensing networks in smart cities, enabling effective noise pollution management and urban planning.
\end{abstract}
 
\IEEEpeerreviewmaketitle
\section{Introduction}
\IEEEPARstart{N}{oise} pollution, also known as environmental noise or sound pollution, refers to unwanted or excessive sound that disrupts human activities and negatively impacts human health \cite{7}. The known sources of noise pollution include transportation (such as road traffic), industrial activities, construction, and urban crowding \cite{8}. Unlike other forms of pollution, noise pollution is invisible and, as a result, is often overlooked despite its severe consequences on public health. Prolonged exposure to high noise levels can lead to hearing loss, cardiovascular diseases, sleep disturbances, and increased stress levels \cite{9}.

Given the widespread impact of noise pollution, continuous and accurate monitoring is essential. Noise pollution monitoring enables policymakers to implement effective mitigation strategies, such as deploying green buffers, sound-absorbing infrastructure, and enforcing zoning regulations to improve urban living conditions. 
Traditionally, noise levels are measured using stationary sound meters deployed by the Central Pollution Control Board (CPCB) in India \cite{10}. As part of this effort, CPCB established the National Ambient Noise Monitoring Network (NANMN), which operates in seven metropolitan cities with monitors deployed at 35 locations across Delhi, Hyderabad, Kolkata, Mumbai, Lucknow, Bangalore, and Chennai. While this system provides reliable data, it often lacks real-time adaptability and sufficient spatial coverage.

The Internet of Things (IoT) has introduced new possibilities for large-scale noise monitoring \cite{Picaut2020, Liu2020}. The IoT-based noise sensors, integrated with cloud computing and wireless communication networks, enable real-time data collection, analysis, and visualization \cite{misra2021introduction}. These systems allow authorities to identify noise pollution hotspots and track temporal variations more effectively. However, despite the advantages of IoT-based noise monitoring, scalability remains a key challenge. Deploying a dense network of sensors requires efficient data transmission, capital, and long-term power management. Sensor reliability, data accuracy, and integration with existing urban infrastructure further complicate large-scale implementation. Also, there is a need for affordable noise measurement networks due to the limitations of expensive professional-grade equipment.


There has been some work on low-cost IoT-based noise monitoring \cite{Linares2016, Middya2023, Dubey2021, Zhang2020, Vazquez2022, Ali2021}. In \cite{Linares2016}, an IoT-based noise monitoring system has been developed using Raspberry Pi and USB microphones. A pilot deployment is carried out near a construction site. In \cite{Middya2023}, a smartphone app named “HonkSense” was developed, which can detect if the data recorded by the smartphone includes honking. The data can be anonymously shared with city planners so they can monitor honking events in the city using maps and timelines. The work in \cite{Dubey2021} uses a crowdsourced data collection approach to map the noise generated during Diwali night in Lucknow. The noise data from 100 locations were collected using a mobile app and then interpolated to predict noise levels at 750 points across the city. The work in \cite{Zhang2020} focuses on predicting environmental noise levels using deep learning. They deployed an IoT-based sensor system to collect noise data in real-time and used this large dataset to train and test the Long Short-Term Memory (LSTM) model. In \cite{Vazquez2022}, a LoRa-based noise monitoring network powered by plant-microbial fuel cells was proposed, using a Kalman filter for dynamic power management. The system enabled battery-free, self-sustained operation with real-time noise estimation in Merida city. In \cite{Ali2021}, a low-cost sensor node with LoRaWAN connectivity was developed to monitor multiple pollutants, which uses ANN-based calibration to improve accuracy, supporting scalable and energy-efficient deployments.

\begin{figure*}[t!]
    \centering
    \begin{subfigure}{0.28\textwidth}
        \centering
        \includegraphics[width=\textwidth]{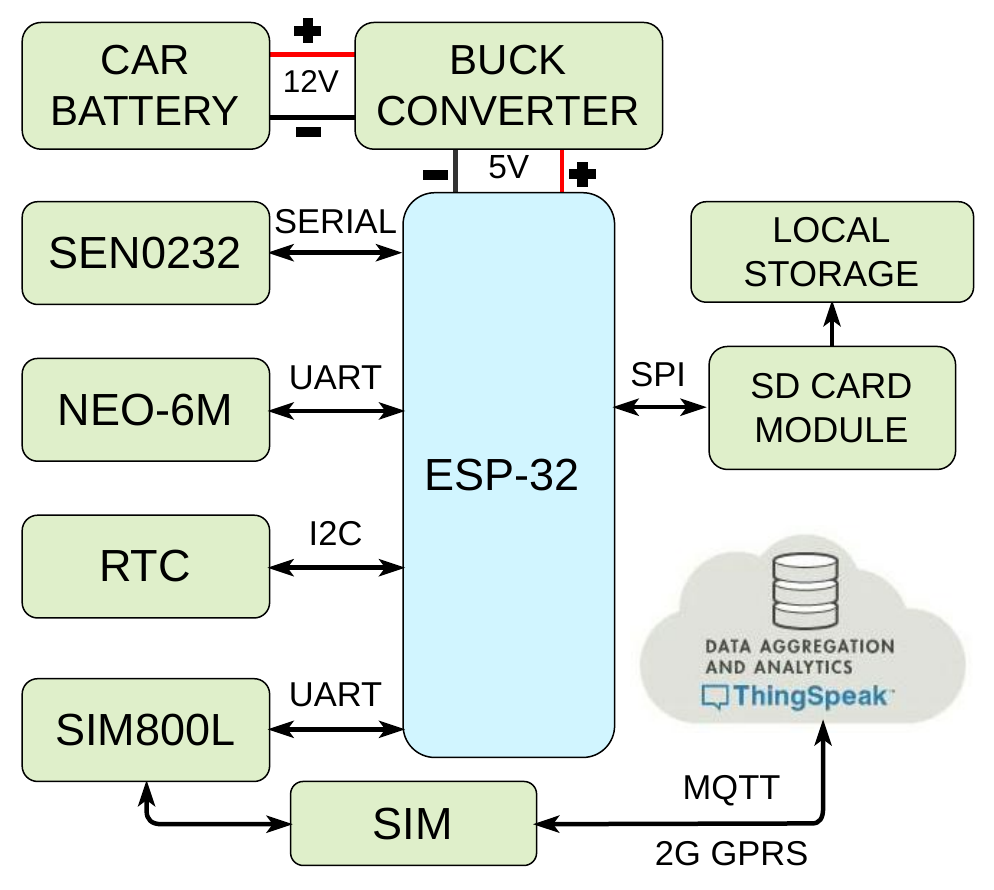}
        \caption{Block architecture}
        \label{fig:Block_diagram}
    \end{subfigure}
    \begin{subfigure}{0.31\textwidth}
        \centering   
        \includegraphics[ width=\textwidth]{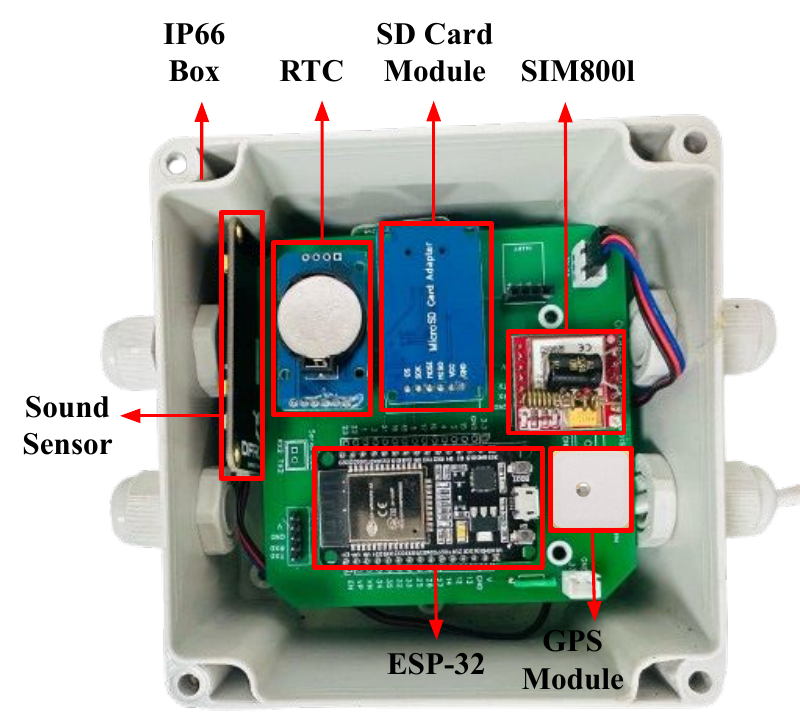}
        \caption{Sound monitoring circuit board}
        \label{fig:Sound_Node}
    \end{subfigure}
    \begin{subfigure}{0.39\textwidth}
    \centering
    \includegraphics[width = 6.5 cm, height = 4 cm]{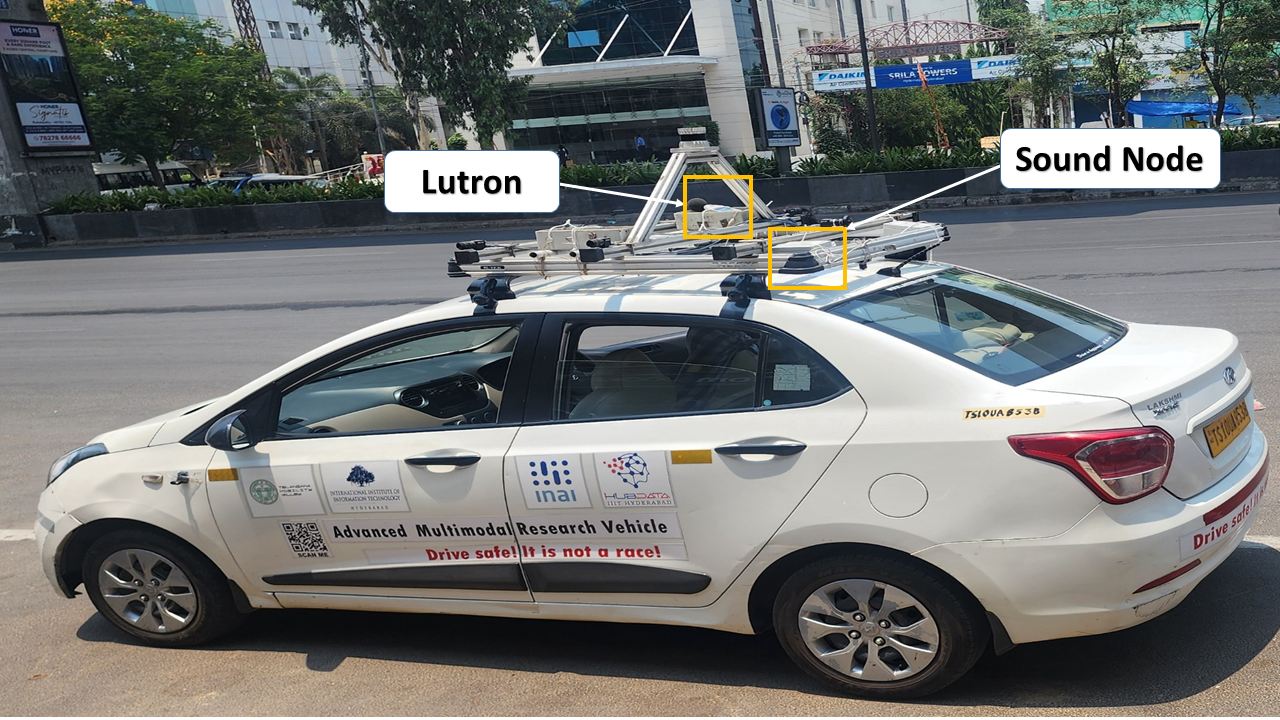} 
    \caption{Actual car setup}
    \label{fig:Car_setup}
        \end{subfigure}
        \caption{Block architecture, circuit board, and the actual car setup of the noise monitoring device.}
        \label{fig:HW}

\end{figure*}

Specific contributions of this paper are 
\begin{itemize}
    \item A low-cost, portable, IoT-based solution is developed for noise pollution monitoring that can be mounted on vehicles, enabling affordable and scalable deployment in urban areas.
    \item For mitigating errors of the low-cost device, they are calibrated against a reference device using machine learning (ML) algorithms in the laboratory under stationary conditions. 
    \item Although the laboratory calibration results are good in the laboratory settings, it is demonstrated that the laboratory calibration is insufficient for the IoT-based noise monitoring devices in the mobile (or \emph{non-stationary}) environment. Next, the calibration is done in a real-world environment while keeping the device and the reference device on the vehicle platform.
    \item Three measurement campaigns\footnote{The data will be made publicly available upon publication to support similar research efforts and enable other researchers to save time and resources.} are carried out in the Indian city of Hyderabad, Telangana State's capital, creating a vast dataset of 4,36,420 points over 27 days over several months. In the first measurement campaign (MC-I), data were collected along a route for over 14 days. In the second measurement campaign (MC-II), data were collected along three routes (different from MC-I) to understand the noise variations across the city. In the third campaign (MC-III), data were collected during Diwali, one of the most popular festivals in India, when many firecrackers burst, elevating the noise levels.
    \item Different kinds of analysis are conducted on the noise data collected in these campaigns: hot-spot detection, temporal and spatial variations, weekday vs weekend variations, and Diwali vs typical day variations.
    \item The effect of traffic on the noise values is demonstrated by correlating the velocity of the measurement vehicle and the noise values. Note that the lower velocity indicates more traffic and congestion, while a higher velocity indicates less traffic.
\end{itemize}
 Unlike previous studies \cite{Linares2016, Middya2023, Zhang2020, Vazquez2022}, which either primarily focused on fixed monitoring solutions or are based on crowdsourcing approaches, our approach offers a practical and scalable alternative. By leveraging low-cost sensor (LCS) and mobile (non-stationary)-based data collection, we enable a broader deployment of noise pollution monitoring in urban areas. Further, our research introduces unique contributions, such as analyzing weekday-weekend variations and festival-specific noise impact, as well as the effect of traffic, which have been largely overlooked in prior works. None of the papers in \cite{Linares2016, Middya2023, Zhang2020, Vazquez2022, Ali2021}, except \cite {Dubey2021}, have data collected in the Indian urban scenario. The work in \cite{Dubey2021} focuses only on one event, Diwali, based on data collected using smartphones on one night. The study in \cite{Ali2021} presents a low-cost, solar-powered air pollution sensor with LoRaWAN and ANN-based calibration but does not focus on noise monitoring or mobile deployments.


\section{Hardware Architecture} \label{hard}
Fig. \ref{fig:HW} shows the hardware block architecture, circuit board for the sound monitoring device (sound node) developed in this work, and the actual node deployed on top of the car. The hardware architecture consists of an ESP32 microcontroller and sensors DFRobot analog sound sensor (SEN0232) for sound levels in dBA, a GPS sensor (NEO-6M) to get the latitude and longitude, a communication GSM module (SIM800L and SIM), a real-time clock module (DS3231 RTC) for timekeeping, a  Secure Digital (SD) card module. The specifications of the individual hardware components are listed in Table I. All the components are connected to the ESP32 microcontroller, where the SEN0232 sensor is connected to the controller through serial communication. The NEO-6M and SIM800L modules are connected to the controller through the UART protocol, the SD card module is connected to the controller through the SPI protocol, and the RTC is connected through the I2C protocol. The sound level meter output is in dBA as the A weighting is intended to simulate the response of a nominal human ear and is also considered by many regulations in many countries to be the best weighting to predict hearing loss due to exposure to noise \cite{IEC61672-1}.

As shown in Fig. \ref{fig:Block_diagram}, the ESP32 microcontroller reads data from the sensor and offloads it to ThingSpeak, a cloud-based server, through the GSM module; it also stores the data in the SD card using the SD card module as a backup to reduce any data loss. The node is configured to record timestamps, sound levels, latitude, and longitude at a 1 s interval, allowing for the monitoring of rapid changes in sound levels.

The device is powered with a 5 V supply derived from a 12 V DC battery in the car through a buck converter and enclosed in an IP66 box made of ABS filament. The enclosure shown in Fig. \ref{fig:Sound_Node} offers complete protection against dust and good protection against water. The form dimensions are width = 125 mm, depth = 125 mm, and height = 125 mm. The node's dimensions are compact enough to comfortably place the node on the top of the car as shown in Fig. \ref{fig:Car_setup}. The overall cost of the device (in terms of bill of material) is Rs. 6000 (approximately 69 USD)\footnote{Assuming the conversion rate of 1 USD = Rs. 87.28 in April 2025.}. 

\renewcommand{\arraystretch}{1.5} 

\begin{table}[t]
    \centering    
    \caption{Hardware component specifications}
    \resizebox{0.49\textwidth}{!}{ 
    \begin{tabular}{|c|c|c|}
        \hline
        \textbf{Component} & \textbf{Specification} & \textbf{Value} \\ 
        \hline
        \multirow{5}{*}{SEN0232 \cite{ref_url_3}} & Operating Voltage & 3.3V - 5V \\ 
        & Measuring Range & 30 dBA - 130 dBA \\ 
        & Frequency Response & 100 Hz - 10 kHz \\ 
        & Output Signal & Analog Voltage Output \\ 
        & Operating Temperature & -40°C to 85°C \\ 
        \hline
        \multirow{5}{*}{NEO-6M \cite{ref_url_4}} & Operating Voltage & 2.7V - 3.6V \\ 
        & Positioning Accuracy & 2.5m (CEP50) \\ 
        & Update Rate & 1 Hz (default), up to 5 Hz \\ 
        & Protocols Supported & NMEA, UBX, RTCM \\ 
        & Operating Temperature & -40°C to 85°C \\ 
        \hline
        \multirow{5}{*}{SIM800L \cite{ref_url_5}} & Operating Voltage & 3.4V - 4.4V \\ 
        & Network Support & 2G (GPRS) \\ 
        & Baud Rate & 1200 - 115200 bps \\ 
        & Communication Protocol & UART \\ 
        & Operating Temperature & -40°C to 85°C \\ 
        \hline
        \multirow{5}{*}{DS3231 \cite{ref_url_6}} & Operating Voltage & 2.3V - 5.5V \\ 
        & Timekeeping Accuracy & ±2ppm at 0°C to 40°C \\ 
        & Communication Interface & I2C \\ 
        & Battery Backup & Yes (Coin Cell CR2032) \\ 
        & Operating Temperature & -40°C to 85°C \\ 
        \hline
        \multirow{4}{*}{\shortstack{SD Card\\Module \cite{ref_url_7}}} & Operating Voltage & 3.3V - 5V \\ 
        & Communication Protocol & SPI \\ 
        & Storage Capacity & Up to 32GB (FAT32 Format) \\ 
        & Operating Temperature & -25°C to 85°C \\ 
        \hline
        \multirow{5}{*}{ESP32 \cite{ref_url_8}} & Operating Voltage & 3.3V \\ 
        & Max Clock Frequency & 240 MHz \\ 
        & Wireless Connectivity & Wi-Fi 802.11 b/g/n, Bluetooth BLE \\ 
        & Communication Interfaces & UART, SPI, I2C \\ 
        & Operating Temperature & -40°C to 85°C \\ 
        \hline
    \end{tabular}
    \label{tab:specifications}
    }
\end{table}

\begin{figure}[tbhp]
    \centering
    \begin{subfigure}{0.21\textwidth}
        \centering
        \includegraphics[width=\textwidth]{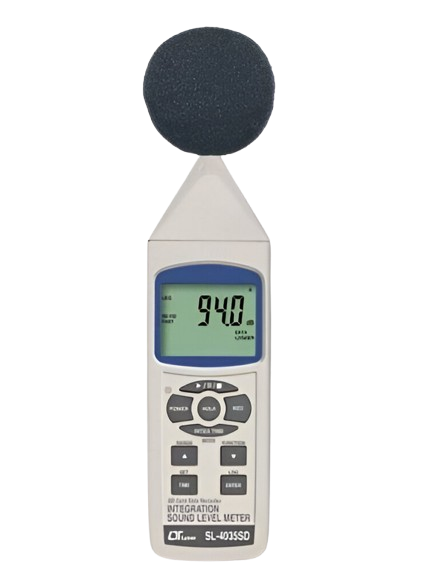}
        \caption{Lutron}
        \label{fig:Lutron}
    \end{subfigure}
    \begin{subfigure}{0.16\textwidth}
        \centering
        \includegraphics[width=\textwidth]{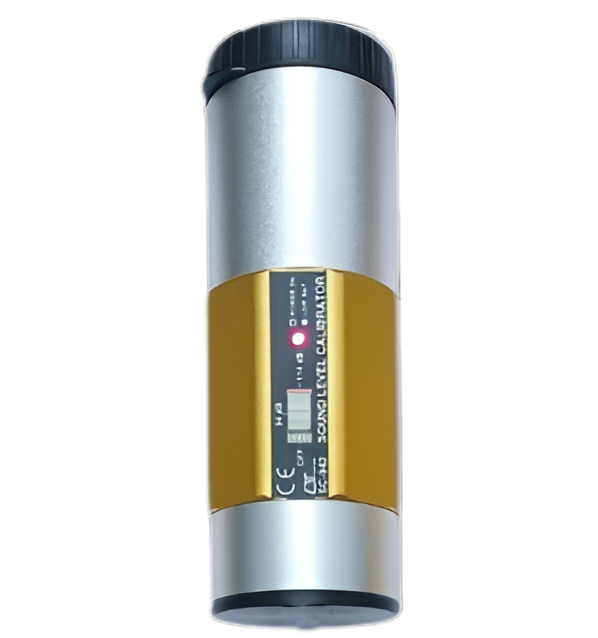} 
        \caption{Calibrator}
        \label{fig:Calibrator}
    \end{subfigure}
    \caption{Lutron (reference instrument) and the calibrator for the reference instrument.}
    \label{fig:Calibration}
\end{figure}
\subsection{Reference instrument} \label{reff}
Fig. \ref{fig:Lutron} illustrates the Lutron SL-4033SD \cite{ref_url_1}, a portable sound level meter designed for precise noise measurement and data logging. It features an electret condenser microphone and operates within a frequency range of 31.5 Hz to 16 kHz, covering a broad spectrum of sound levels from 30 to 130 dB with a resolution of 0.1 dB, with output available in A and C frequency weighting. The device supports customizable sampling intervals, ranging from 1 s to 1 h, with logged data stored in CSV format on an SD card for easy retrieval and analysis. 

To ensure measurement accuracy, the Lutron SL-4033SD is calibrated using the Lutron SC-942 \cite{ref_url_2} Electronic Sound Calibrator shown in Fig. \ref{fig:Calibrator}, which provides a 94 dB and 114 dB reference signal at 1 kHz. The calibration process involves placing the calibrator over the microphone and selecting the desired calibration level. This ensures the sound level meter maintains high precision in its readings, minimizing measurement errors. With its advanced capabilities, user-friendly design, and regular calibration, the SL-4033SD is widely recognized for its reliability and accuracy in environmental noise monitoring.
\begin{figure*}[t]
    \centering
    \setlength{\fboxrule}{0.01pt} 
    \setlength{\fboxsep}{0.0pt}  

    \begin{subfigure}{0.32\textwidth}
        \centering
        \fbox{\includegraphics[width=6cm]{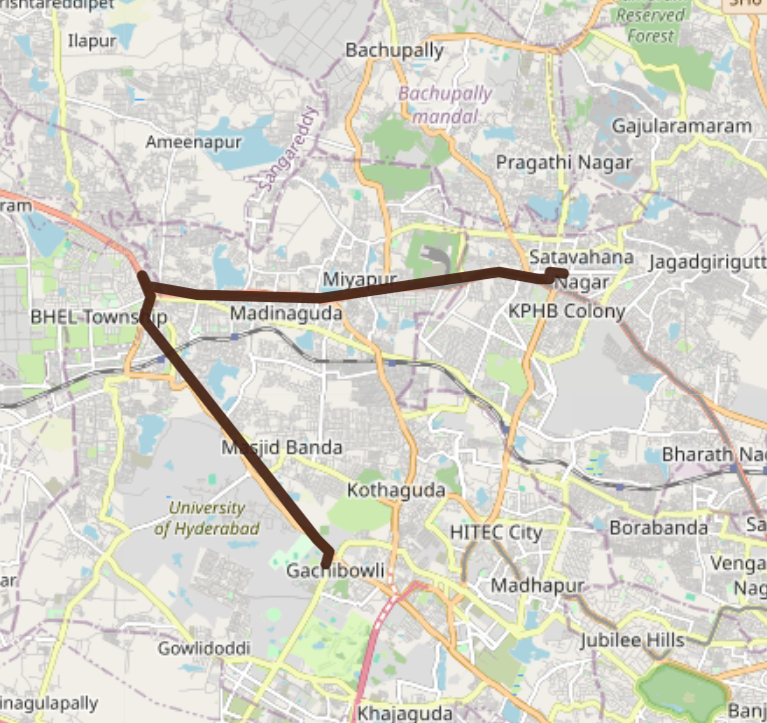}} 
        \caption{MC-I route}
        \label{fig:MC-I}
    \end{subfigure}
    \hfill
    \begin{subfigure}{0.32\textwidth}
        \centering
        \fbox{\includegraphics[width=6cm]{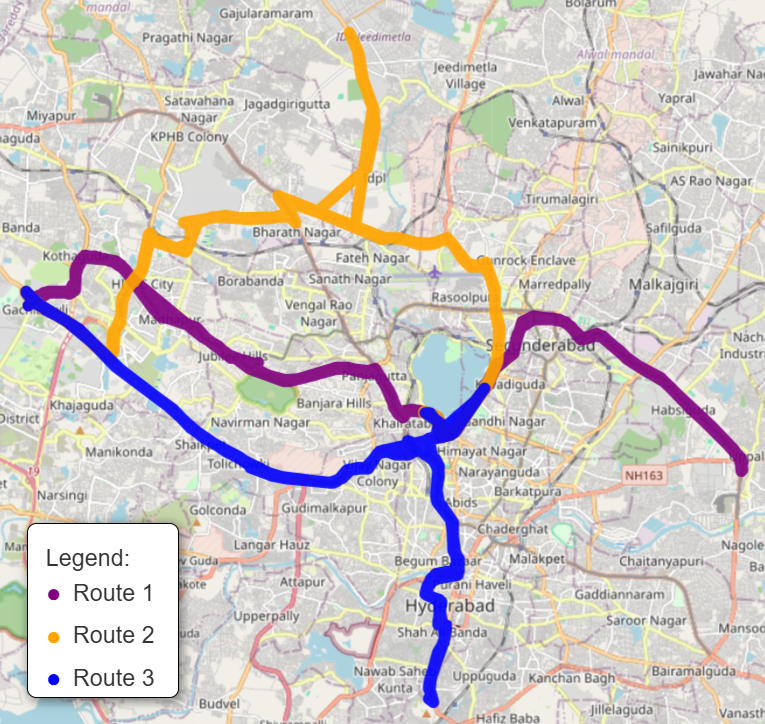}} 
        \caption{MC-II routes}
        \label{fig:MC-II}
    \end{subfigure}
    \hfill
    \begin{subfigure}{0.32\textwidth}
        \centering
        \fbox{\includegraphics[width=6cm]{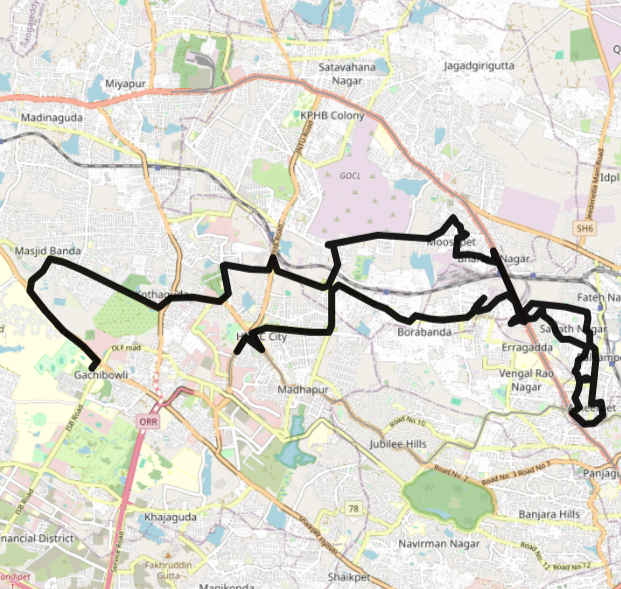}} 
        \caption{MC-III route}
        \label{fig:MC-III}
    \end{subfigure}    

    \caption{Selected routes for measurement campaigns (Best viewed in color).}
    \label{fig:SelectedRoutes}
\end{figure*}

\section{Measurement setup}

 Fig. \ref{fig:Car_setup} illustrates the measurement setup used for the data collection campaign. The setup consists of a reference instrument (Lutron) and a custom sound node, both mounted on the top of a vehicle. A battery powers the Lutron device, while the sound node is powered by a 5 V supply derived from the 12 V battery of the vehicle as specified in section~\ref{hard}. A windshield was placed over the sensors to minimize wind-induced noise interference. 

The measurement campaign was carried out over 27 days, in three phases during which the vehicle tried to maintain speed in the 20–30 km/h range. Given a sampling interval of 1 s, this speed corresponds to a travel distance of approximately 5–8 m between successive measurements, ensuring proper spatial resolution of the noise data.  

Each measurement cycle records the following parameters, which are stored on both an SD card and ThingSpeak:  

\begin{itemize}
    \item \textit{DateTime}: The measurement timestamp, including the date and time, is formatted as dd:mm:yyyy hh:mm:ss.
    \item \textit{Latitude}: Geographic latitude in the WGS84 coordinate system, expressed in decimal degrees.
    \item \textit{Longitude}: Geographic longitude in the WGS84 coordinate system, expressed in decimal degrees.
    \item \textit{Noise level measurement}: Using the sound node in dBA.
    \item \textit{Noise reference level measurement}: Using the reference instrument in dBA to calibrate and cross-check node results.
\end{itemize}
\section{Measurement Campaign}

Data was collected in two experimental settings: a stationary setting inside a laboratory and a non-stationary (mobile) setting on a moving car. Three measurement campaigns were carried out for the mobile setting along the routes shown in Fig. \ref{fig:SelectedRoutes}. 
\subsection{Laboratory data collection} \label{lab}
The IoT-based noise node and the reference
instrument were placed in a closed room, where music was
played at a wide
range of sound values from 50 to 90 dBA. Data was recorded
continuously for over 9 hours with a sampling interval of
one second; a total of 32,461  samples were collected during this period.
\subsection{Measurement Campaign-I (MC-I): Weekends and weekdays} 
MC-I was carried out in Hyderabad along the route shown in Fig. \ref{fig:MC-I} starting from IIIT Hyderabad to Pragati Nagar via BHEL and Miyapur areas of Hyderabad, covering a round-trip distance of around 74 km on each day for over 12 days in May 2024. This route is considered because it contains many nearby hospitals and schools, and is one of the busiest routes in the city.
The data was collected in the morning for three hours (approximately 9 AM - 12 PM) and in the evening for around two and a half hours (approximately 5 PM - 7.30 PM). The focus of this MC was to see the difference in noise levels over weekends and weekdays. Note that four of the twelve days were weekends (Saturday and Sunday), while the other eight days were weekdays. The total data collected in this MC is 1,64,000 points.
      
\begin{table}[b]
    \centering
    \small
    \caption{Noise data collection schedule for MC-II}
    \setlength{\tabcolsep}{5pt} 
    \renewcommand{\arraystretch}{1.2} 
    \resizebox{0.5\textwidth}{!}{ 
    \begin{tabular}{|c|c|c|c|c|c|}
        \hline
        \textbf{Date} & \multicolumn{2}{c|}{\textbf{Morning (9am - 1pm)}} & \multicolumn{2}{c|}{\textbf{Evening (3:30pm - 8pm)}} & \textbf{Route} \\
        \cline{2-5}
        & \textbf{Duration (Hours)} & \textbf{Distance (km)} & \textbf{Duration (Hours)} & \textbf{Distance (km)} &  \\
        \hline
        25th June  & 3.5 & 67 & 4.5 & 67 & 1 \\
        \hline
        26th June  & 4 & 74 & 3.5 & 74 & 2 \\
        \hline
        28th June  & 3.5 & 62 & 4 & 62 & 3 \\
        \hline
        29th June  & 3.5 & 66 & \cellcolor{pink}\shortstack{Rescheduled \\ due to rain} & & 1 \\
        \hline
        1st July   & \cellcolor{pink}\shortstack{Rescheduled \\ due to rain} & & 4 & 67 & 1 \\
        \hline
        2nd July   & 4 & 74 & 3.5 & 74 & 2 \\
        \hline
        3rd July   & 3.5 & 61 & 4 & 60 & 3 \\
        \hline
        4th July   & 3.5 & 67 & 4 & 67 & 1 \\
        \hline
        5th July   & 4 & 74 & 3.5 & 74 & 2 \\
        \hline
        6th July   & \cellcolor{pink}\shortstack{Rescheduled \\ due to rain} & & 4 & 72 & 1 \\
        \hline
        7th July   & 4 & 73 & 3.5 & 74 & 2 \\

        \hline
        8th July   & 4 & 74 & \cellcolor{pink}\shortstack{Rescheduled \\ due to rain} & & 2 \\
        \hline
        9th July   & 3.5 & 62 & 4 & 61 & 3 \\
        \hline
    \end{tabular}
    }
    \label{tab:NoiseSchedule}
\end{table}   
      
\subsection{Measurement Campaign-II (MC-II): Different routes} 
MC-II was carried out in Hyderabad along the three routes shown in Fig. \ref{fig:MC-II}. This phase spanned two weeks and focused on collecting live traffic noise data during morning and evening hours on three selected routes. Table \ref{tab:NoiseSchedule} specifies a detailed data collection schedule.

\begin{itemize}
    \item \textit{Route 1}: IIIT Hyderabad to Uppal via Madhapur and Hussain Sagar, covering a round-trip distance of approximately 67 km with a travel time of around 3.5 hours.
    \item \textit{Route 2}: IIIT Hyderabad to Jeedimetla via Hussain Sagar, spanning approximately 74 km round-trip for around 4 hours.
    \item \textit{Route 3}: IIIT Hyderabad to Falaknuma Palace, covering a round-trip distance of about 61 km with a travel time of around 3.5 hours.
\end{itemize}

The routes were carefully chosen based on high commuter density to capture significant noise variations. Data was collected during the morning and evening periods to ensure a complete analysis of traffic-related noise patterns. The data was collected for around 13 days during June-July 2024 with the detailed schedule specified in Table \ref{tab:NoiseSchedule}. In this measurement campaign, approximately 2,42,492 data points were collected.

\subsection{Measurement Campaign-III (MC-III): Diwali} 

This MC focused on understanding the variation of noise levels during Diwali compared to a typical day. MC-III was carried out in Hyderabad along the route shown in Fig. \ref{fig:MC-III} starting from IIIT Hyderabad to the interiors of SR Nagar, Sanath Nagar to cover sample residential areas where festival celebrations are significant. Data was collected on October 31, 2024 (Diwali) and November 6, 2024 (typical day) for around 4.5 hours, covering a round-trip distance of around 70 km daily. In this measurement campaign, over 29,928 data points were collected.

\section{Data preprocessing and calibration}
\subsection{Pre-processing} \label{preproc}
\begin{figure}[b]
    \centering
    \includegraphics[width = 1\columnwidth]{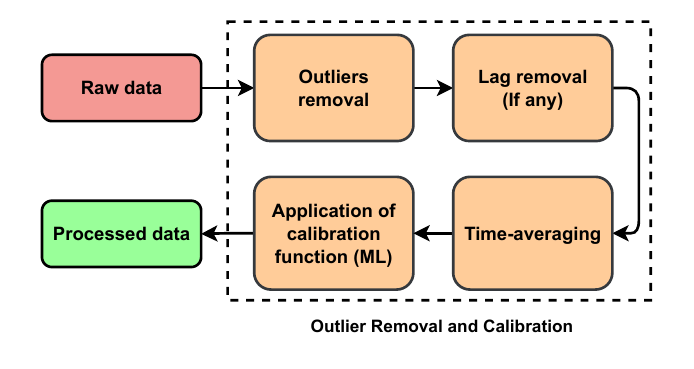}
    \caption{Data preprocessing and calibration.}
    \label{fig:data_preprocessing}
\end{figure}

The data collected from the sound node undergoes preprocessing before applying trained ML models, as illustrated in Fig. \ref{fig:data_preprocessing}.  

\begin{itemize}
    \item \textit{Outliers removal:} Out of the total data points of 4,36,420 from three MCs, 2,444 outliers, representing 0.56\% of the data, were removed using the Interquartile Range (IQR) method. As IQR ensures robust outlier detection in dynamic environments \cite{boukerche2020outlier}, the approach is particularly suitable for noise measurements, where sudden variations due to honking or fireworks occur.   
    \item \textit{Lag removal:} A slight time lag between the reference instrument and the sound node, caused by differences in data logging, was corrected using cross-correlation analysis. The lag corresponding to the highest correlation was identified and applied to align the signals. Pearson's correlation was calculated to assess similarity, while Mean Absolute Error (MAE) and Root Mean Squared Error (RMSE) were used to validate alignment accuracy. This correction accounts for logging or transmission delays, ensuring accurate comparisons.  
    \item \textit{Time-averaging:} To improve correlation between sound node data and reference device data before calibration, the data were time-averaged to 10 s intervals. This time interval corresponds to a spatial distance of approximately 100 m, ensuring that essential variations in noise levels are preserved while reducing short-term fluctuations.  
\end{itemize}
 
\subsection{Calibration} \label{calib_methods}
At the output of sensors, sound level indicators must be adjusted to give measures as close as possible to sound levels measured by a reference instrument, such as an acoustic calibrator. In the best case, this adjustment considers the variability of the microphone sensitivity. Still, this adjustment can also correct linearity defects in the acquisition system or the digital processing chain. In the literature, two calibration methods are considered \cite{segura2014low}. The first calibration method uses an acoustic calibrator (mainly 94 dB at 1000 Hz), which is only possible when the microphone mounting device allows it (matching the diameter of the microphone mounting with the acoustic calibrator). The second calibration method compares the data from the low-cost node with a reference sound level meter under the same measuring conditions \cite{hakala2010design}. In this paper, the first method is used for calibrating the reference instrument as explained in the subsection \ref{reff}, and the second method was used to calibrate the IoT-node against the reference instrument as explained in the subsection \ref{lab}. For the second method, since the target data from the reference instrument was available, supervised learning \cite{Sarker2021} was used to calibrate the IoT-based sound node by aligning its measurements with a reference sensor's. The preprocessed noise data from the sound node served as input features, while the reference sensor data acted as the target output. ML-based regression models were used to establish a mapping between these variables and to minimize discrepancies between predicted and actual values. A 10-fold cross-validation approach was implemented for laboratory and mobile calibration processes. The following regression algorithms were applied in this study \cite{Sarker2021}:
\begin{itemize}
    \item \textit{Simple Linear Regression (SLR)}: SLR is a widely used technique for modelling the relationship between a dependent variable and one or more independent variables. In SLR, the relationship is represented by
    \begin{equation}
    Y = a + bX + e,
    \label{eq:simple_linear}
    \end{equation}
    where \( Y \) is the dependent variable, \( X \) is the independent variable, \( a \) is the intercept, \( b \) is the slope, and \( e \) is the error term. This equation defines a straight line that best fits the data by minimizing the difference between predicted and actual values. 
     \item \textit{Multiple Linear Regression (MLR)}: MLR can be used when there is more than one dependent variable. MLR extends the SLR approach by incorporating multiple independent variables, expressed as
    \begin{equation}
    Y = a + b_1X_1 + b_2X_2 + ... + b_nX_n + e,
   \label{eq:multiple_linear}
   \end{equation}
   where \( X_1, X_2, ..., X_n \) are different predictor variables. This method allows for a more accurate representation of complex relationships in data, making it a useful tool for prediction and analysis.
   However, SLR is most appropriate for the sound node in static measurements since only one input variable exists. However, for mobile (moving) measurements, we tried MLR by taking the vehicle's velocity as one variable.
    \item \textit{Segmented Regression (SR)}: Segmented regression is a piecewise linear modelling technique for different linear relationships in distinct data ranges. It is useful when the data exhibits varying slopes beyond a specific breakpoint. The general form of SR can be expressed as
    \begin{equation}
    Y =
    \begin{cases} 
    a_1 + b_1 X + e_1, & \text{if } X < X_b, \\
    a_2 + b_2 X + e_2, & \text{if } X \geq X_b,
    \end{cases}
    \label{eq:segmented_regression}
    \end{equation}
     where \( X_b \) is the breakpoint, \( (a_1, b_1) \) and \( (a_2, b_2) \) are the intercept and slope for each segment, and \( e_1, e_2 \) are the error terms.
     This work applies SR to noise data, where preprocessed noise levels are used as the predictor variable, and reference instrument values are the target variable. The dataset is split at an optimal breakpoint, determined dynamically, allowing different linear models to capture distinct trends in the data. This method ensures accurate modelling of complex noise variations.
    \item \textit{Polynomial Regression (PR)}: PR is an extension of linear regression that models the relationship between an independent variable \( X \) and a dependent variable \( Y \) using an \( n \)th-degree polynomial. This method is useful when data follows a nonlinear pattern. The general equation for PR is
    \begin{equation}
    Y = a + b_1X + b_2X^2 + b_3X^3 + ... + b_nX^n + e,
    \label{eq:polynomial}
    \end{equation}
    where \( a \) is the intercept, \( b_1, b_2, ..., b_n \) are regression coefficients. When data does not follow a linear trend but exhibits a curved pattern, PR provides a better fit by capturing higher-order relationships.
    \item \textit{Support Vector Regression (SVR)}: SVR is a supervised learning algorithm for predicting continuous values. It extends the Support Vector Machines (SVM) concept for regression tasks by identifying a hyperplane that best fits the data while allowing a margin of tolerance. The SVR model aims to minimize the error while maintaining generalization. The SVR model predicts the dependent variable \( Y \) (reference values) using an independent variable \( X \) (shifted noise level), defined as
    \begin{equation}
     Y = f(X) + e,
     \label{eq:svr}
     \end{equation}
    where \( f(X) \) represents the function learned by the model. A radial basis function (RBF) kernel is commonly used to map the input space into a higher-dimensional feature space, enabling the model to capture complex relationships. In this work, the SVR model is trained using preprocessed noise levels as the input feature and the reference instrument values as the target variable. The model evaluates the relationship between sound node measurements and the reference instrument, enabling effective noise measurement.

    \item \textit{Decision Tree (DT)}: DTs are widely used non-parametric supervised learning models for classification and regression tasks. A DT consists of nodes, branches, and leaf nodes, where each internal node represents a decision based on an attribute, and each leaf node provides the final prediction. The model sorts data from the root node down to a leaf node by applying a series of binary rules. For regression tasks, the final prediction is determined by averaging the values of the dependent variable within the leaf node. While decision trees effectively capture complex relationships, they are prone to overfitting, mainly when the tree depth is not controlled.

    \item \textit{Random Forest Regression (RFR)}: RFR is an ensemble learning method that builds multiple decision trees and combines their predictions to improve accuracy and reduce overfitting. It follows a ``parallel ensembling''
 approach, where several decision trees are trained on different subsets of the data in parallel. The final prediction is obtained by averaging (for regression) or majority voting (for classification). 
 
To introduce variation among the trees, RFR uses two key techniques: 
\begin{enumerate}[label=(\roman*)]
    \item Bootstrap aggregation (Bagging): Each tree is trained on a randomly sampled subset of the data (with replacement), ensuring diversity in training.
    \item Random features: Instead of considering all features at each split, RFR selects a random subset, which prevents dominance by highly correlated features and improves generalization.
\end{enumerate}

The RFR model is adaptable to classification and regression tasks, making it suitable for various applications. It typically achieves higher accuracy than individual decision trees by reducing variance and mitigating overfitting.

\end{itemize}

\begin{figure*}[tbh]
    \centering
    \begin{subfigure}{0.49\textwidth}
        \includegraphics[width=\linewidth]{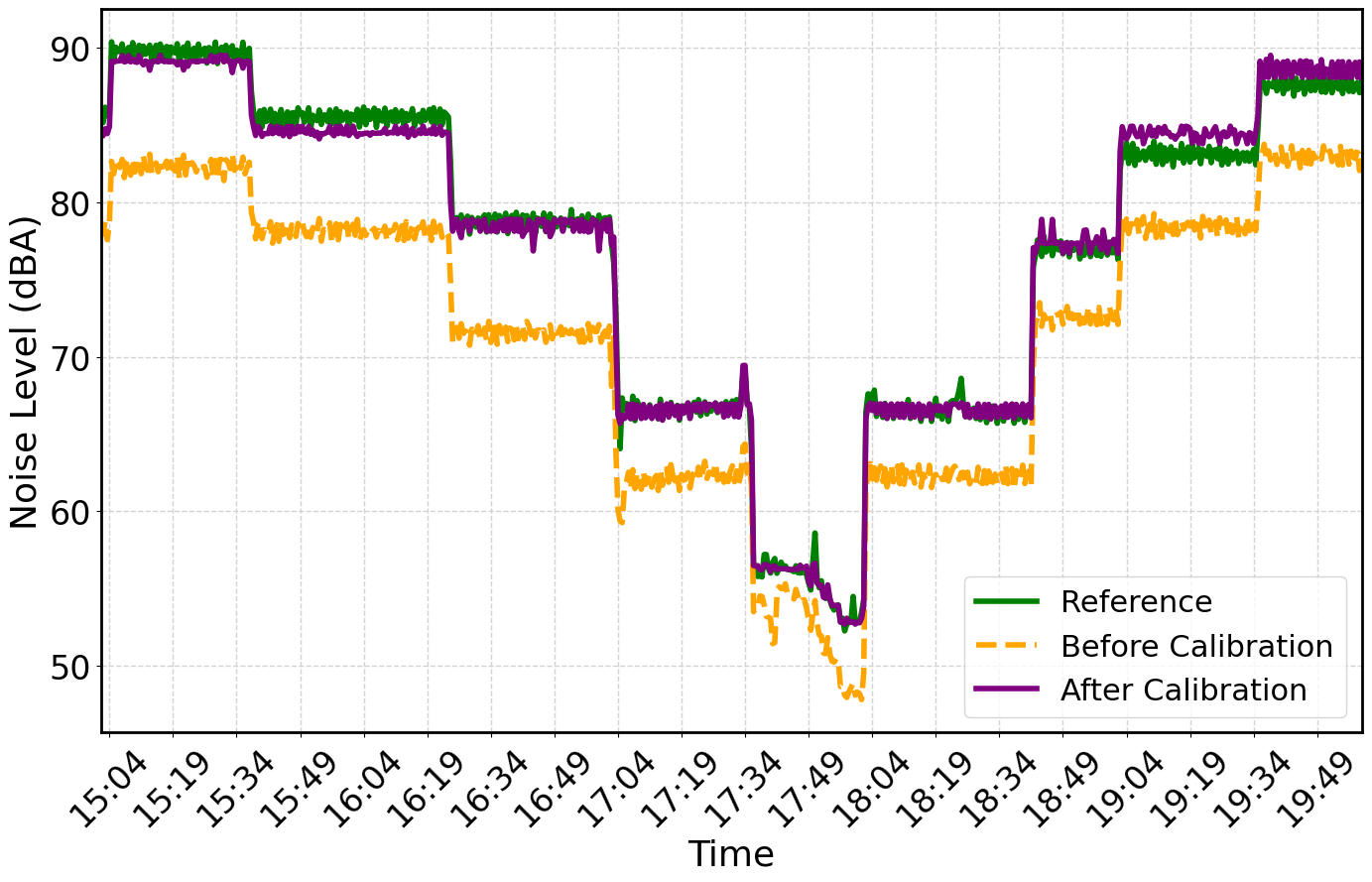}
        \caption{Laboratory calibration using RFR}
        \label{fig:image1}
    \end{subfigure}
    \hfill
    \begin{subfigure}{0.49\textwidth}
        \includegraphics[width=\linewidth]{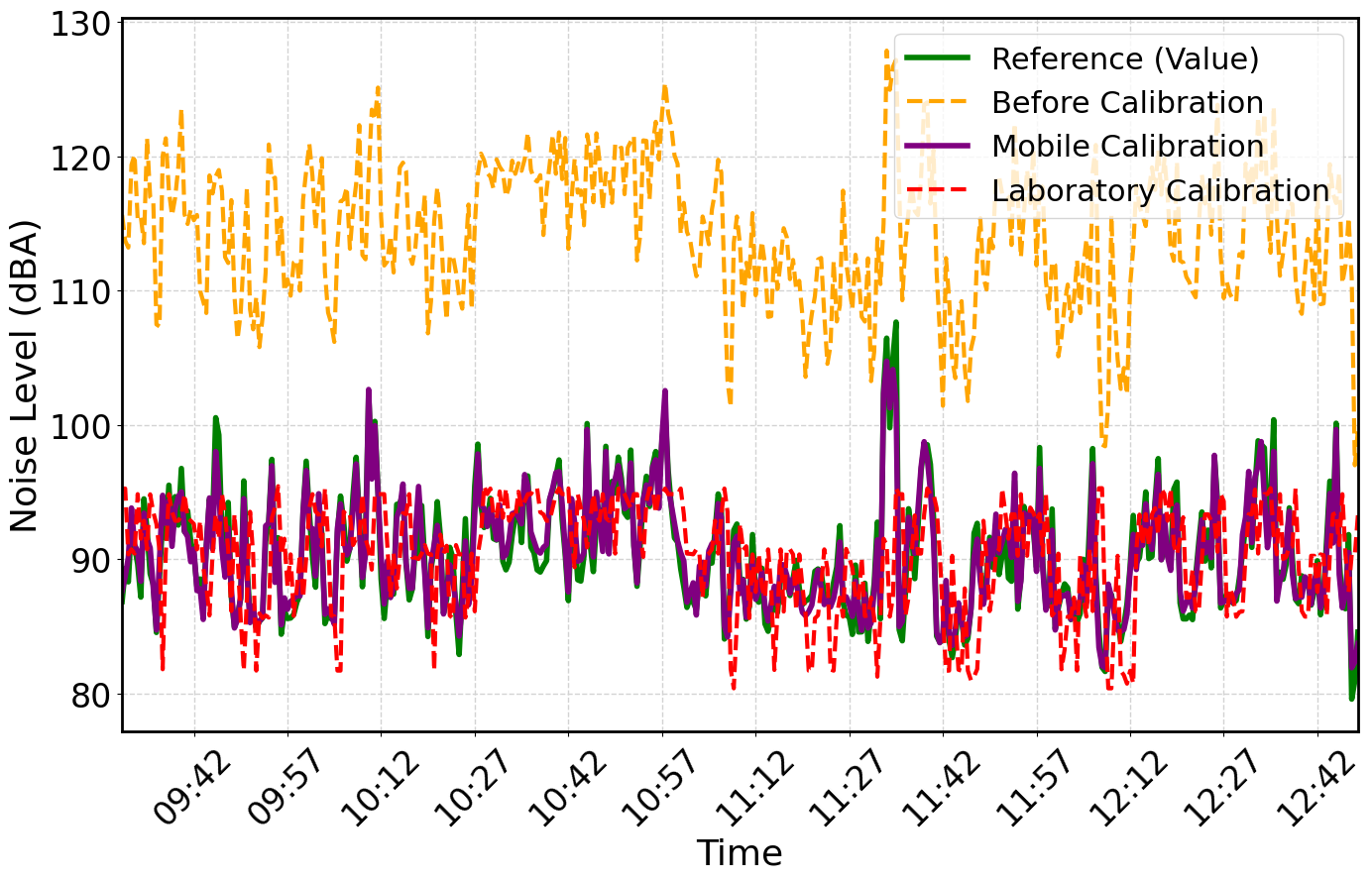}
        \caption{Laboratory and mobile calibration on mobile data using RFR}
        \label{fig:image2}
    \end{subfigure}
    \caption{Time series plots of laboratory and mobile calibration results (Best viewed in color).}
    \label{fig:combined}
\end{figure*}

\subsection{Performance evaluation}

The performance of the calibration algorithms is evaluated using three key metrics \cite{willmott2005advantages}: \( \mathrm{R}^2 \), MAE, and RMSE. These performance indicators are described as follows:

\begin{itemize}
    \item \textit{Coefficient of Determination ( $\mathrm{R}^2$ )}: This metric quantifies the proportion of variance in the dependent variable that is explained by the independent variable(s) in a regression model. It indicates the reduction in prediction error, with a value closer to 1 signifying a better fit. The \( \mathrm{R}^2 \) value is computed as \cite{R2}
    \begin{equation}
    \textrm{R}^2 = 1 - \frac{\mathrm{SS}_{\mathrm{res}}}{\mathrm{SS}_{\mathrm{tot}}},
    \label{eq:r_squared}
    \end{equation}
    where $\mathrm{SS}_{\mathrm{res}}$ and  $\textrm{SS}_{\textrm{tot}}$ represent the sum of squared residuals, and the total sum of squares, respectively and are given by   
\begin{eqnarray}
\mathrm{SS}_{\mathrm{tot}} &=& \sum_{i=1}^{n} (\mathrm{y}_i - \bar{\mathrm{y}})^2, 
    \label{eq:sstot} \\
    \mathrm{SS}_{\mathrm{res}} &=& \sum_{i=1}^{n} (\mathrm{y}_i - \hat{\mathrm{y}}_i)^2,
    \label{eq:ssres}
\end{eqnarray}
    where \(n\) is the number of samples, \(\bar{\mathrm{y}}\) is the mean of the observed values $y_i$, and \(\hat{\mathrm{y}}_i\) represents the predicted values.
    \item \textit{Mean Absolute Error (MAE)}: MAE measures the average absolute difference between actual and predicted values, indicating model accuracy. It is calculated as
 
\begin{equation}
    \text{MAE} = \frac{1}{n} \sum_{i=1}^{n} |\mathrm{y}_i - \hat{\mathrm{y}}_i|.
    \label{eq:mae}
    \end{equation}
    \item \textit{Root Mean Squared Error (RMSE)}: RMSE represents the standard deviation of the residuals and provides insight into model performance by emphasizing larger errors. It is defined as
    \begin{equation}
    \text{RMSE} = \sqrt{\frac{1}{n} \sum_{i=1}^{n} (\mathrm{y}_i - \hat{\mathrm{y}}_i)^2}.
    \label{eq:rmse}
    \end{equation}
    We also evaluated the correlation coefficient and p-value for null hypothesis testing.
    \item \textit{Correlation Coefficient}: This metric quantifies the strength and direction of the linear relationship between actual and predicted values. It is computed as
   \begin{equation}
   {r} = \frac{\sum_{i=1}^{n} (\mathrm{y}_i - \bar{\mathrm{y}})(\hat{\mathrm{y}}_i - \bar{\hat{\mathrm{y}}})}
    {\sqrt{\sum_{i=1}^{n} (\mathrm{y}_i - \bar{\mathrm{y}})^2} \sqrt{\sum_{i=1}^{n} (\hat{\mathrm{y}}_i - \bar{\hat{\mathrm{y}}})^2}},
    \label{eq:correlation}
    \end{equation}
    where \(\bar{\mathrm{y}}\) and \(\bar{\hat{\mathrm{y}}}\) represent the mean of actual and predicted values, respectively.   
    \item \textit{P-value of the correlation coefficient}: This metric evaluates the statistical significance of the observed Pearson correlation coefficient \( r \). To test the null hypothesis that there is no linear relationship between the actual and predicted values, the following test statistic is used
\begin{equation}
    t = \frac{r \sqrt{n - 2}}{\sqrt{1 - r^2}},
    \label{eq:t_statistic}
\end{equation}
where \( n \) is the number of paired observations. Under the null hypothesis, the test statistic follows a t-distribution with \(n-2 \) degrees of freedom. The corresponding two-tailed p-value is then computed, indicating the probability of observing such a correlation by random chance. A p-value below a conventional threshold (e.g., 0.05) indicates that the correlation is statistically significant \cite{di2020statistical}.
\end{itemize}

\section{Calibration results} 
To evaluate the effectiveness of the calibration approaches discussed earlier, we applied them to both laboratory and mobile data. This section presents the comparative performance of various ML models in different sensing environments, starting with laboratory calibration.

\subsection{ML-based laboratory calibration}
\begin{table}[tbhp]
    \centering
    \small
    \caption{Performance comparison of different laboratory ML models on laboratory data}
    \resizebox{0.45\textwidth}{!}{ 
    \begin{tabular}{|l|c|c|c|c|}
        \hline
        \textbf{Method} & \textbf{R$^2$} & \textbf{MAE} & \textbf{RMSE} \\ 
        \hline
        Raw data from node & 0.590 & 18.7 & 21.82 \\ 
        \hline
        Data after pre-processing & 0.977 & 16.80 & 17.92 \\ 
        \hline
        SLR & 0.979 & 1.54 & 2.04 \\ 
        \hline
        PR (order = 4) & 0.978 & 1.55 & 2.06 \\ 
        \hline
        SR & 0.979 & 1.54 & 2.04 \\ 
        \hline
        SVR & 0.977 & 1.53 & 2.12 \\ 
        \hline
        DT (best depth = 3) & 0.983 & 1.37 & 1.89 \\ 
        \hline
        RFR (best depth = 3) & \textbf{0.985} & \textbf{1.33} & \textbf{1.81} \\ 
        \hline
    \end{tabular}
    }
    \label{tab:comparison}
\end{table}

Table~\ref{tab:comparison} shows the performance of the low-cost IoT node for noise monitoring in laboratory conditions. It can be seen from the table that the performance of the raw data is poor with low $\textrm{R}^2=0.59$, large MAE $=18.7$, and RMSE $=21.82$ values given the use of the low-cost sensor. Data preprocessing improves $\textrm{R}^2 = 0.977$, indicating that the preprocessed data follows the trend well with the reference sensor after preprocessing. However, the error is still large in terms of MAE $=16.8$ and RMSE $=17.92$, as there is significant bias, as seen in Fig. 5(a). This demonstrates that these low-cost sensors need calibration.

Table~\ref{tab:comparison} shows the performance of different ML-based calibration techniques described in Section \ref{calib_methods}, which are applied to preprocessed data from the IoT node. It can be seen that all methods demonstrate satisfactory performance with $\mathrm{R}^2$ values exceeding 0.976, and MAE and RMSE as low as 1.33 and 1.81, respectively. 
Fig.~\ref {fig:image1} illustrates the laboratory calibration results for RFR. We can observe that the sound node data follows the trend of the reference instrument with minimal error after calibration.

\subsection{Applying laboratory ML models on the noise data from mobile MCs}
\begin{table}[tbp]
    \centering
    \small    \caption{Performance comparison of different laboratory data ML models on mobile data}
    \resizebox{0.45\textwidth}{!}{ 
    \begin{tabular}{|l|c|c|c|c|}
        \hline
        \textbf{Method} & \textbf{R$^2$} & \textbf{MAE} & \textbf{RMSE} \\ 
        \hline
        Raw data from node & 0.16 & 23.94 & 24.87 \\ 
        \hline
        Data after pre-processing & 0.69 & 23.86 & 24.07 \\ 
        \hline
        SLR & \textbf{0.63} & \textbf{2.47} & \textbf{3.06} \\ 
        \hline
        PR (order = 4)& 0.57 & 2.57 & 3.33 \\ 
        \hline
        SR & 0.357 & 3.57 & 3.10 \\ 
        \hline
        SVR & 0.581 & 2.68 & 3.29 \\ 
        \hline
        DT (best depth = 3) & 0.571 & 2.73 & 3.33 \\ 
        \hline
        RFR (best depth = 3) & 0.585 & 2.68 & 3.27 \\ 
        \hline
    \end{tabular}
    }
    \label{tab:compar}
\end{table}

Next, ML models from laboratory calibration were applied to data collected during mobile MC-I. However, as shown in Table~\ref{tab:compar}, the performance significantly degrades as compared to when the same models were applied to lab data. For example, $\mathrm{R}^2$ has fallen in the range of 0.357--0.69, while MAE and RMSE have increased and are in the range 2.47--3.57 and 3.1--3.33, respectively. These poor performance metrics are because of different distributions and variance of training laboratory data and testing mobile data; these findings demonstrate that laboratory calibration models fail to generalize to data, emphasizing the need for alternative approaches for mobile calibration by using proper training data.

\subsection{ML-based calibration using data from mobile MCs}
\begin{table}[tbhp]
    \centering
    \small
    \caption{Performance comparison of mobile ML models}
    \resizebox{0.45\textwidth}{!}{ 
    \begin{tabular}{|l|c|c|c|c|}
        \hline
        \textbf{Method} & \textbf{R$^2$} & \textbf{MAE} & \textbf{RMSE} \\ 
        \hline
        Raw data from node & 0.160 & 23.94 & 24.87 \\ 
        \hline
        Data after pre-processing & 0.690 & 23.86 & 24.07 \\ 
        \hline
        SLR & 0.739 & 1.68 & 2.23 \\ 
        \hline
        PR (order = 4) & 0.774 & 1.51 & 2.07 \\ 
        \hline
        SR & 0.763 & 1.95 & 2.47 \\ 
        \hline
        SVR & 0.776 & 1.49 & 2.06 \\ 
        \hline
        DT (best depth = 5) & 0.786 & 1.51 & 2.02 \\ 
        \hline
        RFR (best depth = 5) & \textbf{0.937} & \textbf{0.85} & \textbf{1.09} \\ 
        \hline
    \end{tabular}
    }
    \label{tab:compar3}
\end{table}

Next, ML models were trained using MC data, where the IoT node and the reference devices were kept on top of a moving car as discussed in Section~\ref{calib_methods}. Table~\ref{tab:compar3} shows the performance of ML models trained on the mobile data. It can be seen that the performance for all methods has increased. However, RFR performs best with $\mathrm{R}^2$ $= 0.93$, MAE $= 0.85$, and RMSE $= 1.09$.  
Fig. \ref{fig:combined} shows that after calibration, the corrected data closely follow the trend of the reference data. These findings strongly support the need for mobile calibration to use mobile data for real-time noise measurements.

The correlation coefficient is evaluated for all calibration methods, which is satisfactory, and the statistical significance of the correlation was cross-verified using the \textit{p}-value for all calibration methods. In each case, the \textit{p}-value was found to be extremely low ($\ll 0.05$), 
indicating that the observed correlation is highly significant and not due to a random chance.

\section{Application of calibrated mobile sensing}
Given that the mobile ML models perform much better than the laboratory ML models for calibrating the data collected in MCs, only mobile ML models are used in analyzing the results for the three mobile MCs. The results are presented in terms of time series, box plots, and spatial plots for better data visualization. Unless specified otherwise, all results are based on 10 s averaging.

\begin{figure*}[tbhp]
    \centering
    \setlength{\fboxrule}{0.01pt} 
    \setlength{\fboxsep}{0.0pt}  
    \begin{subfigure}{0.45\textwidth}
        \centering        \fbox{\includegraphics[width=\linewidth]{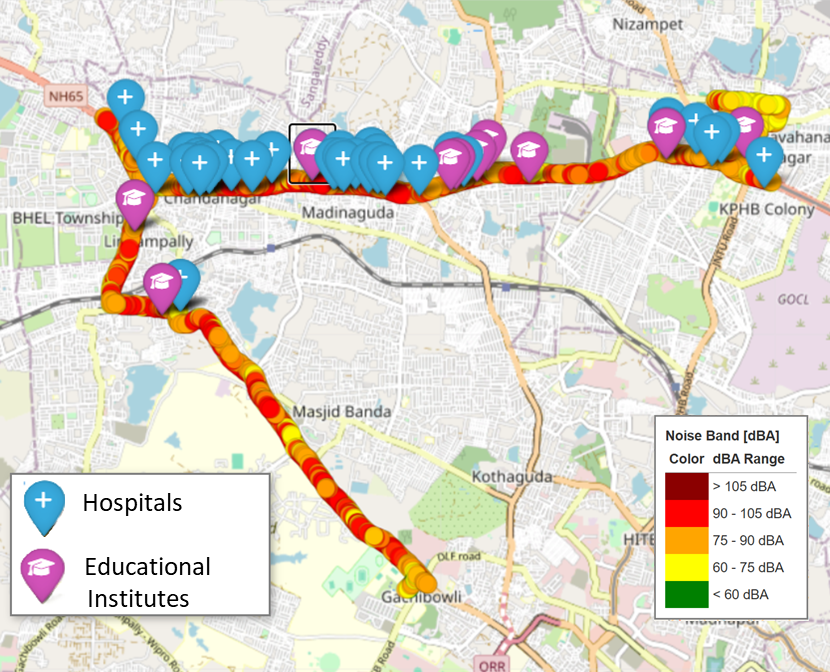}}
        \caption{MC-I route}
        \label{MC1}
    \end{subfigure}
    \hfill
    \begin{subfigure}{0.45\textwidth}
        \centering        \fbox{\includegraphics[width=\linewidth]{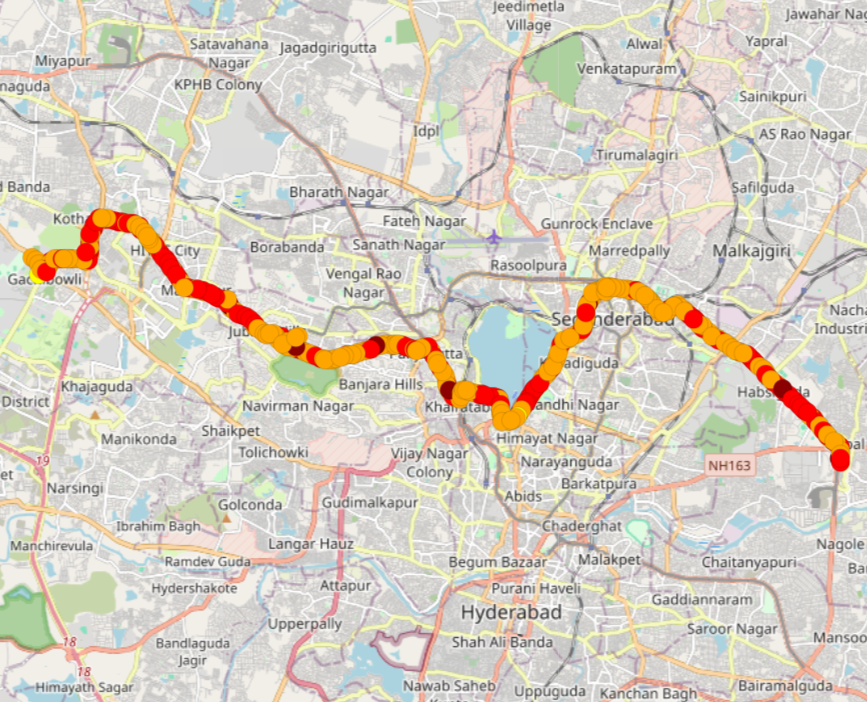}}
        \caption{MC-II route-1}
        \label{MC2-1}
    \end{subfigure}
    \vspace{1cm}
    \begin{subfigure}{0.45\textwidth}
        \centering        \fbox{\includegraphics[width=\linewidth]{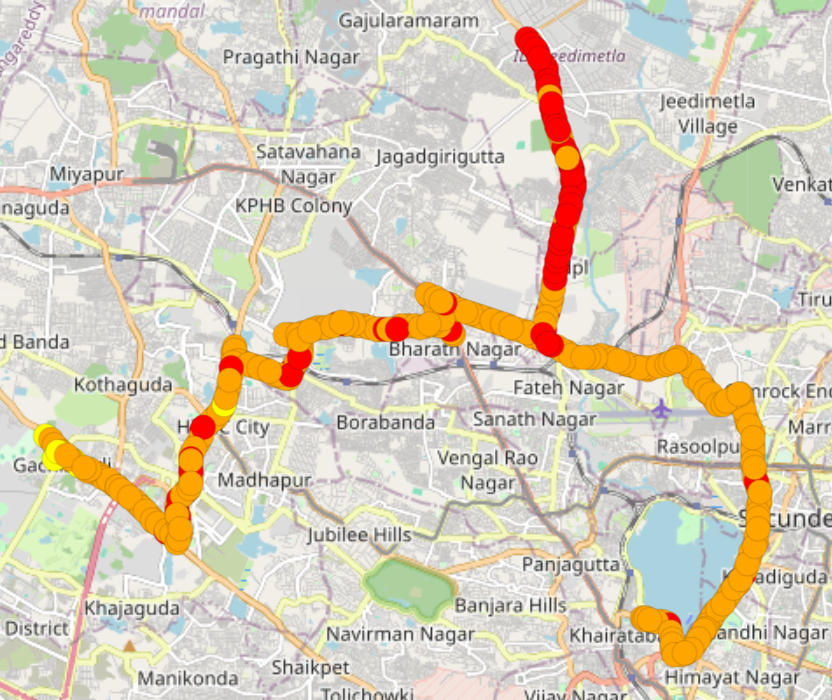}}
        \caption{MC-II route-2}
        \label{MC2-2}
    \end{subfigure}
    \hfill
    \begin{subfigure}{0.45\textwidth}
        \centering        \fbox{\includegraphics[width=\linewidth]{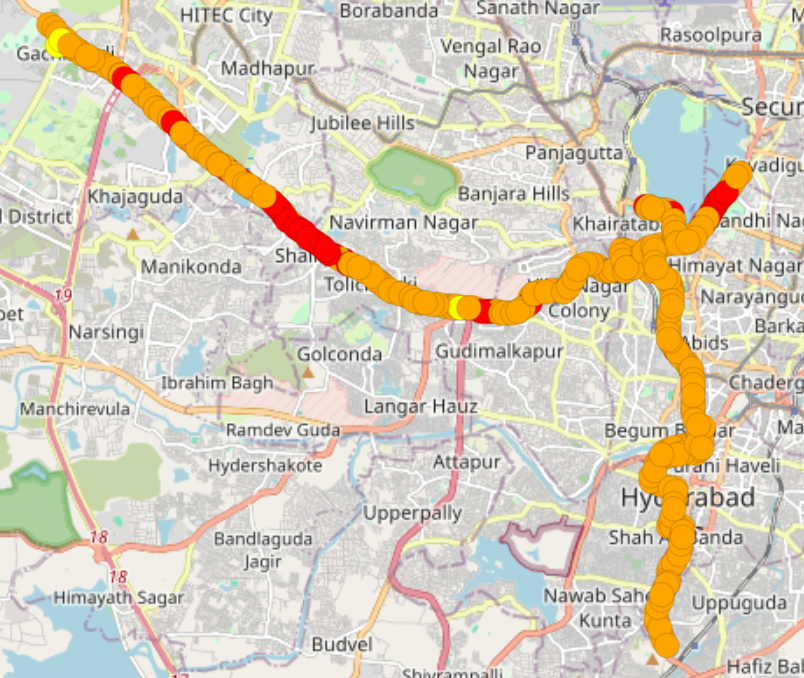}}
        \caption{MC-II route-3}
        \label{MC2-3}
    \end{subfigure}
    \caption{Noise maps of all the selected routes showing spatial variations (Best viewed in color).}
    \label{fig:route_maps}
\end{figure*}

\begin{figure*}[tbhp]  
    \centering
    \begin{subfigure}{0.49\textwidth}
        \centering
        \includegraphics[ width=\textwidth]{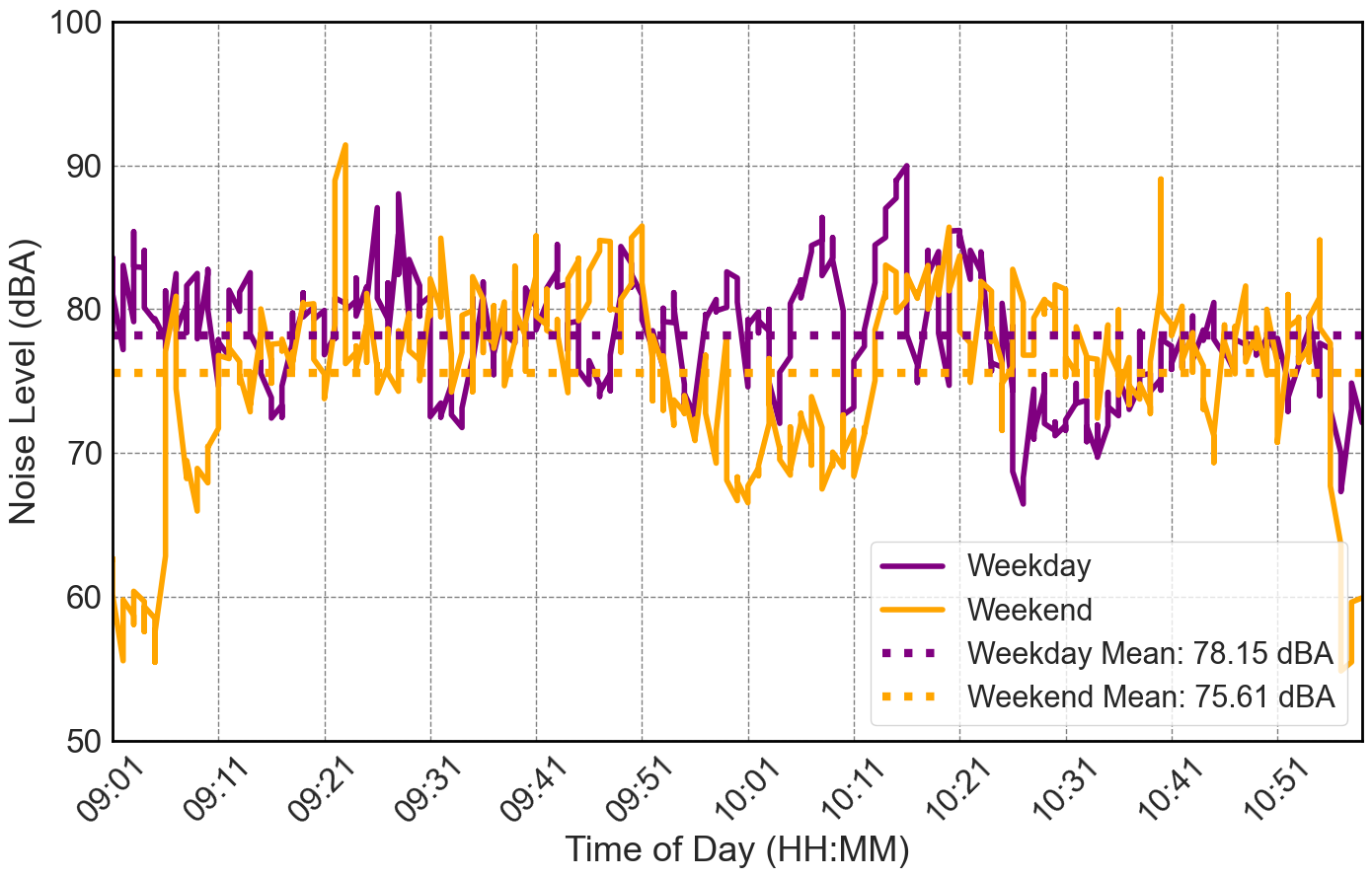}  
        \caption{Morning time-averaged noise levels}
        \label{fig:time_series_1}
    \end{subfigure}
    \hfill
    \begin{subfigure}{0.49\textwidth}
        \centering        \includegraphics[width=\textwidth]{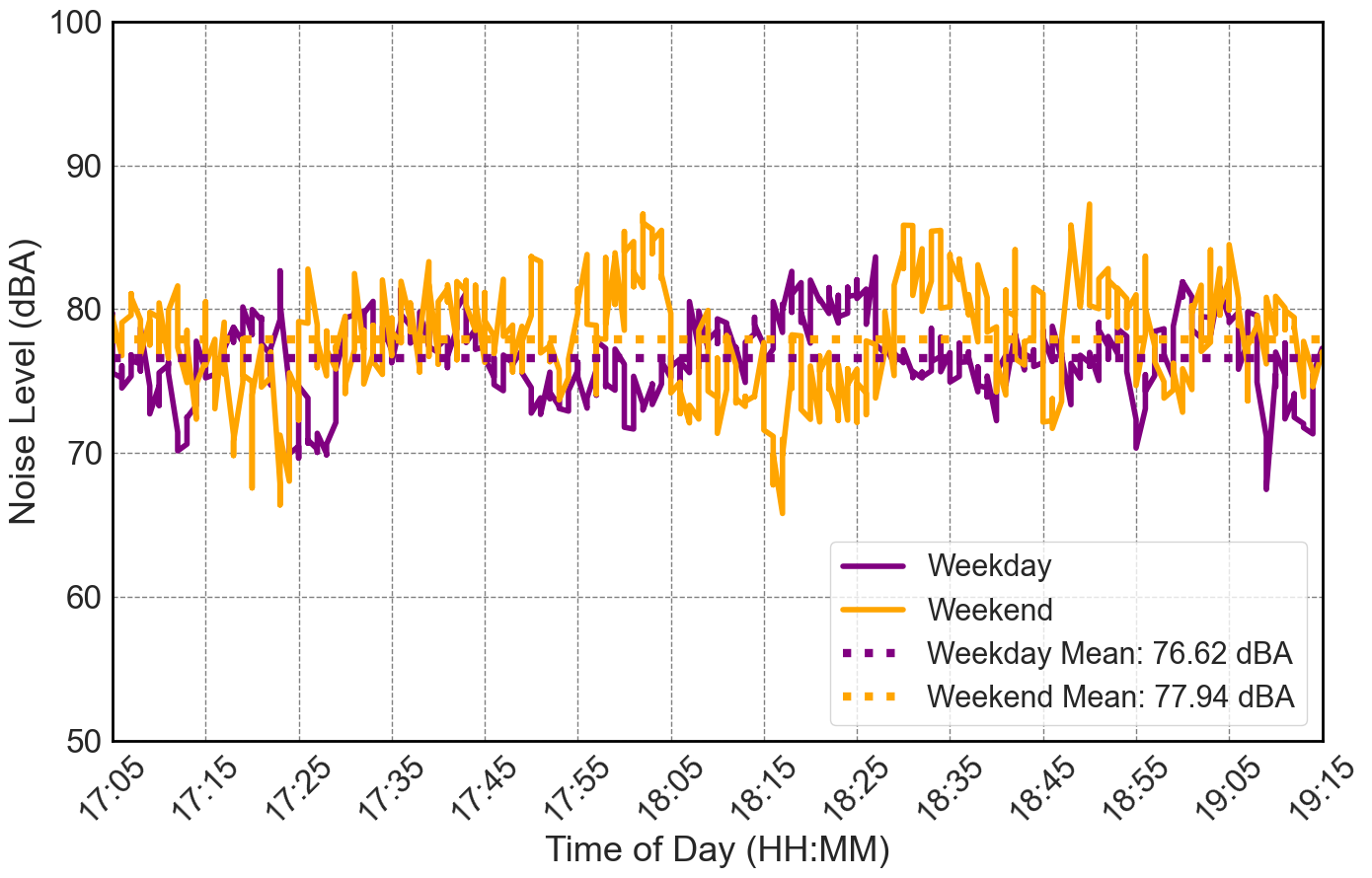}  
        \caption{Evening time-averaged noise levels}
        \label{fig:time_series_2}
    \end{subfigure}
    \hfill
    \caption{Time series plots of the average noise values for the selected MC-I route during the morning and evening, averaged over weekdays and weekends (Best viewed in color).}
    \label{fig:time_series_all}
\end{figure*}

\subsection{MC I}

\begin{figure}[tbhp]
    \centering    \includegraphics[width=\columnwidth]{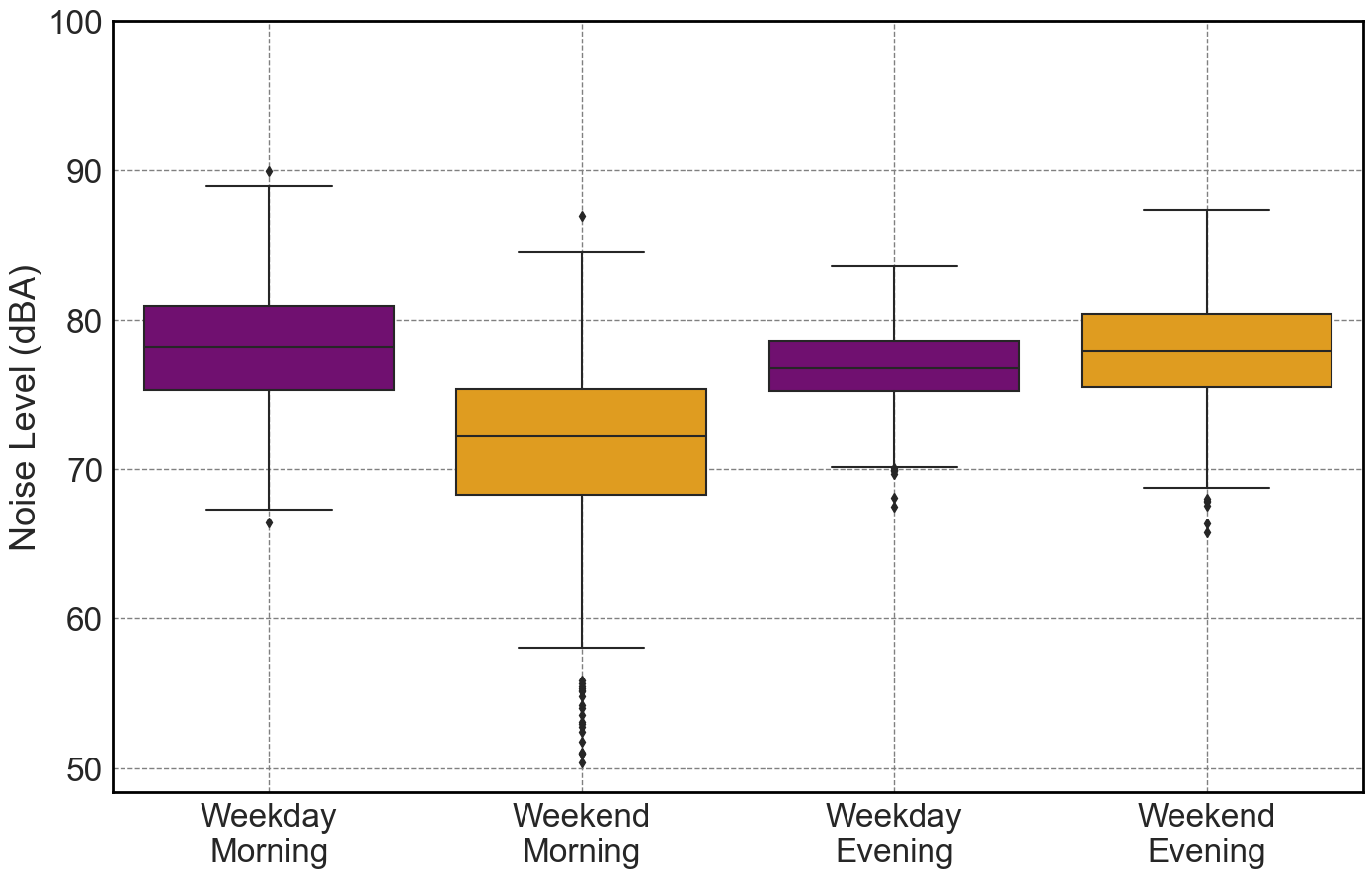}  
    \caption{Box plot of weekday and weekend noise levels for MC-I (Best viewed in color).}
    \label{fig:may box plot}
\end{figure}

Fig. \ref{MC1} shows the spatial variations of noise levels across the selected route of MC-I; those levels are spatially averaged over 12 days of data collection. It can be seen that most of the locations on this route have average values of more than 75 dBA. There are many hospitals and educational institutes along this route, which are marked in the figure. Several stretches of the route adjacent to hospitals and educational institutions had noise levels consistently surpassing the CPCB’s daytime threshold of 50 dBA \cite{noise_rules_2000} prescribed for silence zones, indicating elevated exposure risks for vulnerable populations. 

Fig. \ref{fig:time_series_all} shows the time series plot of the noise values for the selected MC-I route during the morning and evening, averaged over weekdays and weekends, and Fig. \ref{fig:may box plot} shows the box plot of the same. Note that the data collection for these days was done almost during the same time slots. In the morning, the mean noise value over the entire weekday route is 78.32 dBA, 2.4 dBA higher than the mean value over the weekends (75.61 dBA). Also, the variance in the noise values is lower on weekdays than on weekends. This is expected as on weekdays, the traffic patterns are determined by work-related travel, which is time-constrained.
On the other hand, people's weekend plans can change every weekend.  Another observation from the time series plot is that the noise values on a weekend are lower in the early morning hours, as the traffic only increases after 11 AM, given the holidays. During the evenings, the mean noise levels are nearly the same on weekdays and weekends. However, values are slightly higher on weekend evenings, possibly due to increased vehicular movement for shopping and recreational activities.


\subsection{MC-II}
\begin{figure*}[tbhp]  
    \centering
    \begin{subfigure}{0.49\textwidth}
        \centering        \includegraphics[width=\textwidth]{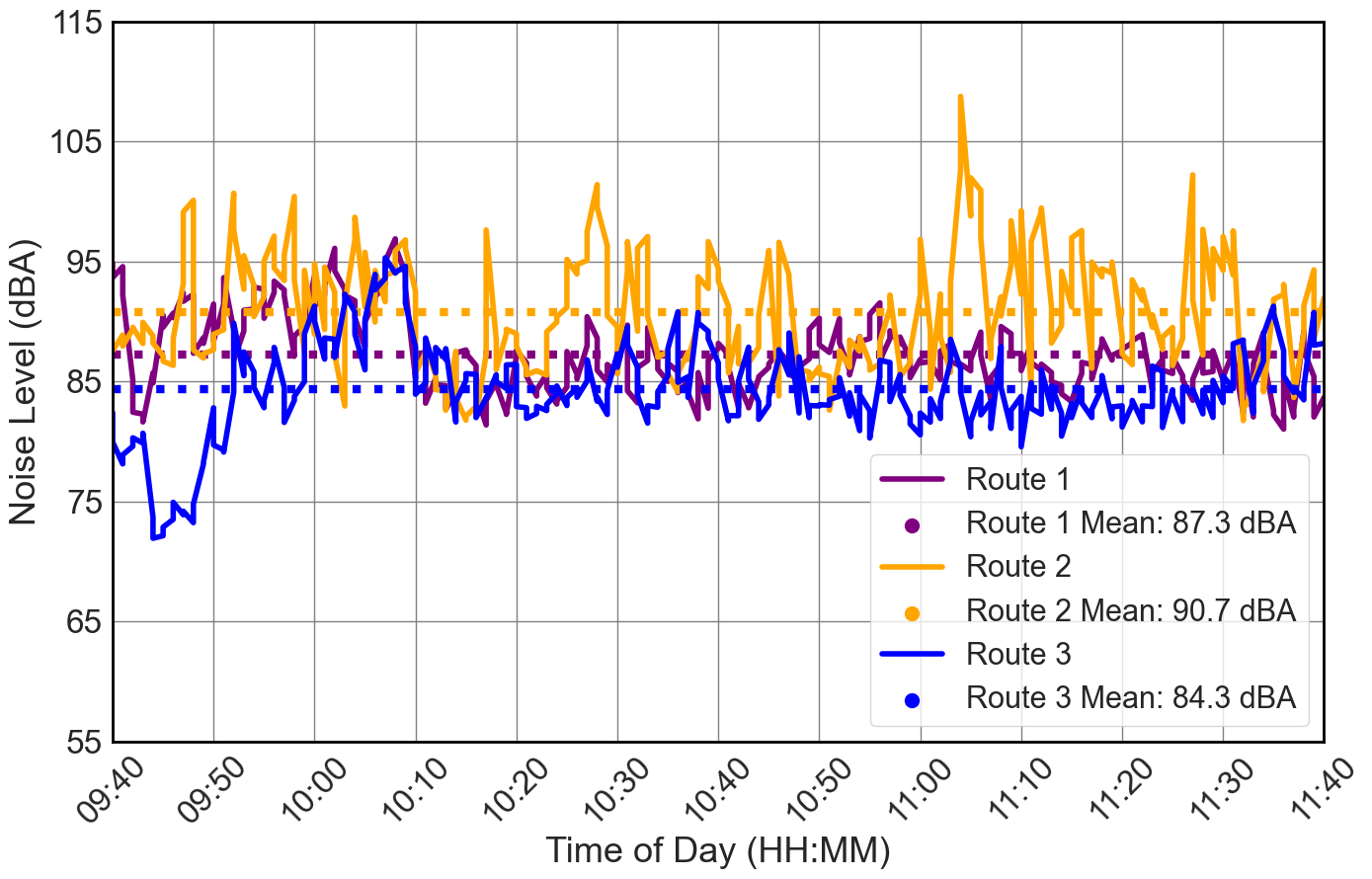}  
        \caption{Morning time-averaged noise levels}
        \label{fig:time_series_1}
    \end{subfigure}
    \hfill
    \begin{subfigure}{0.49\textwidth}
        \centering        \includegraphics[width=\textwidth]{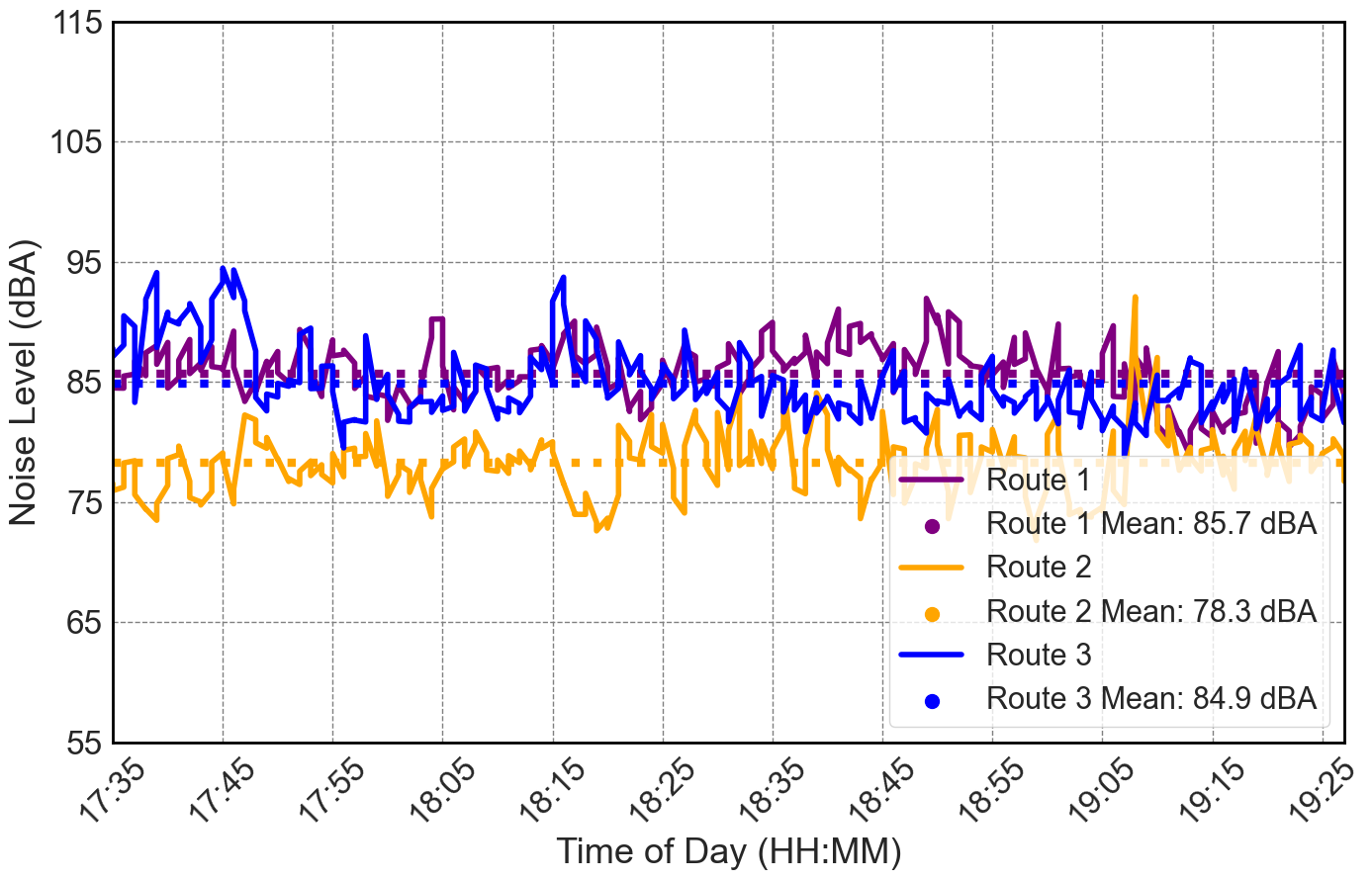}  
        \caption{Evening time-averaged noise levels}
        \label{fig:time_series_2}
    \end{subfigure}
    \hfill
    \caption{Time series plots of noise data comparing morning and evening patterns across three selected routes for MC-II, averaged over the total number of days of data collection on those routes (Best viewed in color).}
    \label{fig:time_series_3_routes}
\end{figure*}

\begin{figure}[tbhp]
    \centering
    \includegraphics[width = 1\columnwidth]{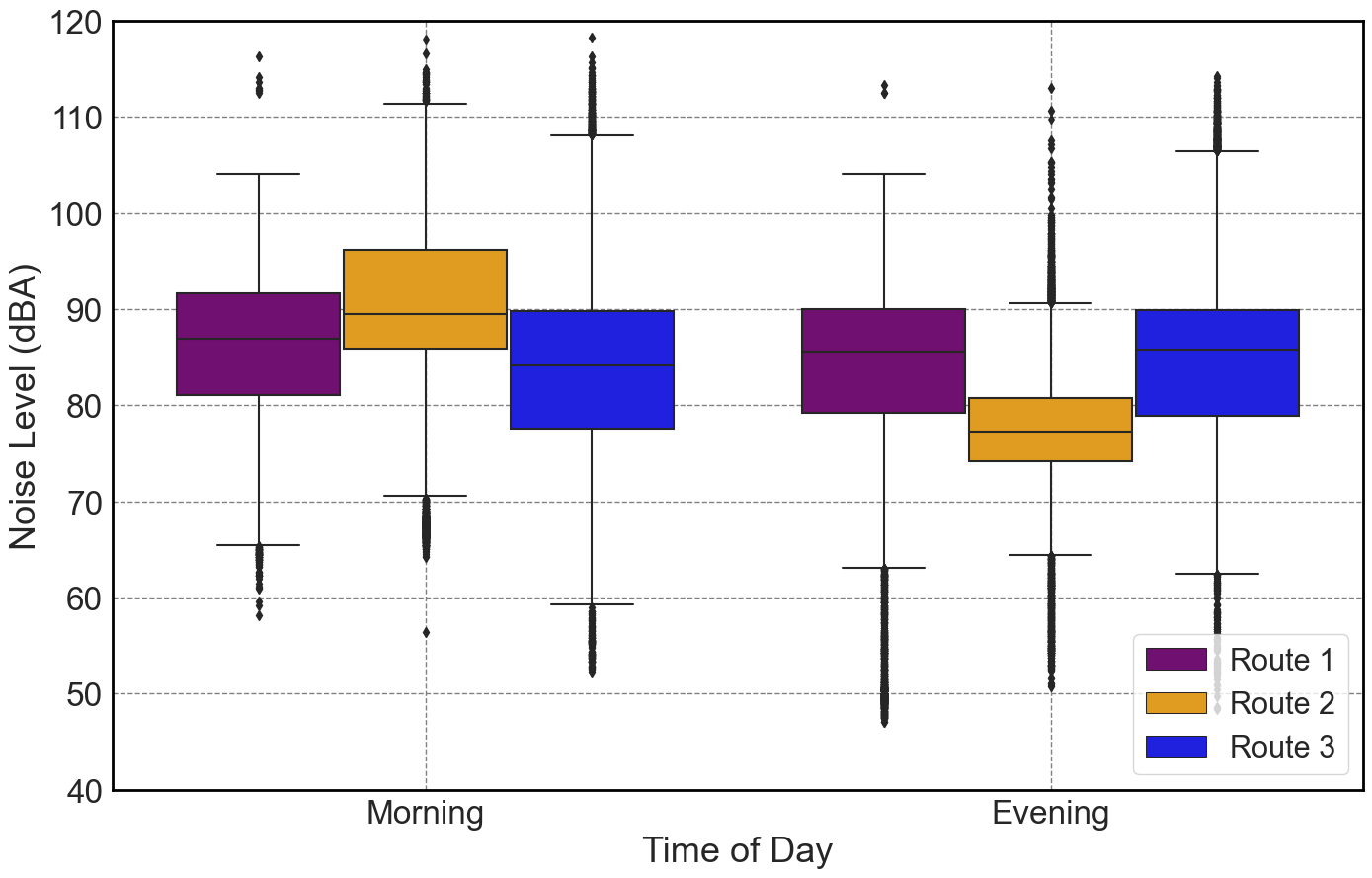}
    \caption{Box plot of morning and evening noise levels for all three selected routes for MC-II (Best viewed in color).}
    \label{fig:all routes box}
\end{figure}

Fig. \ref{MC2-1}--\ref{MC2-3} presents the noise maps for the three routes in MC-II, representing the spatial variation of noise levels on the Hyderabad streets. In these figures, it can be seen that the noise values on the streets of Hyderabad are above 75 dB. Most of the values are in the range 75--90 dBA, which aligns with the data from CPCB on noise values in Indian metropolitan cities \cite{cpcb2021}. 
In most cases, vehicular traffic, including honking, is the major source of noise pollution. 
From these figures, it can be seen that the regions where the noise values are more than 90 dBA are: 
\begin{itemize}
\item Route I: Hitech City, Madhapur, Jubilee Hills, Tank Bund, and Habsiguda. Hitech City and Madhapur have a lot of software companies, which is leading to increased traffic congestion. Jubilee Hills is a prominent upscale residential and commercial area with popular cafes, schools, hospitals, and shopping centers, which is leading to increased local and through traffic to nearby places, Tank Bund is a major tourist attraction near Hussain Sagar Lake, drawing a large number of visitors, especially during weekends and holidays, causing congestion, Habsiguda is a growing residential and educational hub with institutions like Osmania University and nearby research centers, leading to regular traffic from students and professionals. 
\item Route II: Jedimetla, which is an Industrial area. Being a major industrial area with heavy manufacturing units, transport trucks, and machinery operations, it consistently records high noise levels of more than 90dBA. Jeedimetla is known especially for pharmaceuticals, heavy engineering, and fabrication units, which run heavy equipment and freight trucks throughout the day, and employees commute regularly on a shift basis. So, higher noise levels are expected.
\item Route III: Shaikpeth, Tolichowki, and Tank Bund. Shaikpeth is a busy residential area with a high population, leading to increased road and transportation demand. The area is close to several important commercial centers, which results in more vehicles entering and exiting the area throughout the day. Tolichowki is home to many commercial establishments, restaurants, and retail shops, attracting locals and visitors and increasing traffic. It's a major junction connecting areas like Banjara Hills, Gachibowli, and Mehdipatnam, making it a key transit route. Ongoing construction projects for road development or infrastructure improvements will narrow the available lanes, contributing to congestion. Tank Bund is a popular spot for tourists and locals due to its scenic views of Hussain Sagar Lake and nearby landmarks. This increases visitors, especially during weekends and holidays, and it's also near other tourist attractions like NTR Gardens.
\end{itemize}

 Fig. \ref{fig:time_series_3_routes} presents the time series plot of noise values for the three routes averaged over different instances of data collection on those routes. Fig. \ref{fig:all routes box} shows the box plot for the same during morning and evening slots. Both figures corroborate the observations of Figs. \ref{MC2-1}--\ref{MC2-3}, showing that the mean value of noise levels on the routes is more than 75 dBA. Route 2 exhibits the highest average noise levels in the morning, with a mean of approximately 90 dBA among the three routes. In contrast, Route 1 and Route 3 show similar noise levels, averaging around 85 dBA, with minimal variation between morning and evening. However, Route 2 displays greater variation between morning and evening, with evening noise levels approximately 10 dBA lower than morning.

\begin{figure*}[tbhp]  
    \centering
    \begin{subfigure}{0.49\textwidth}
        \centering        \includegraphics[width=\textwidth]{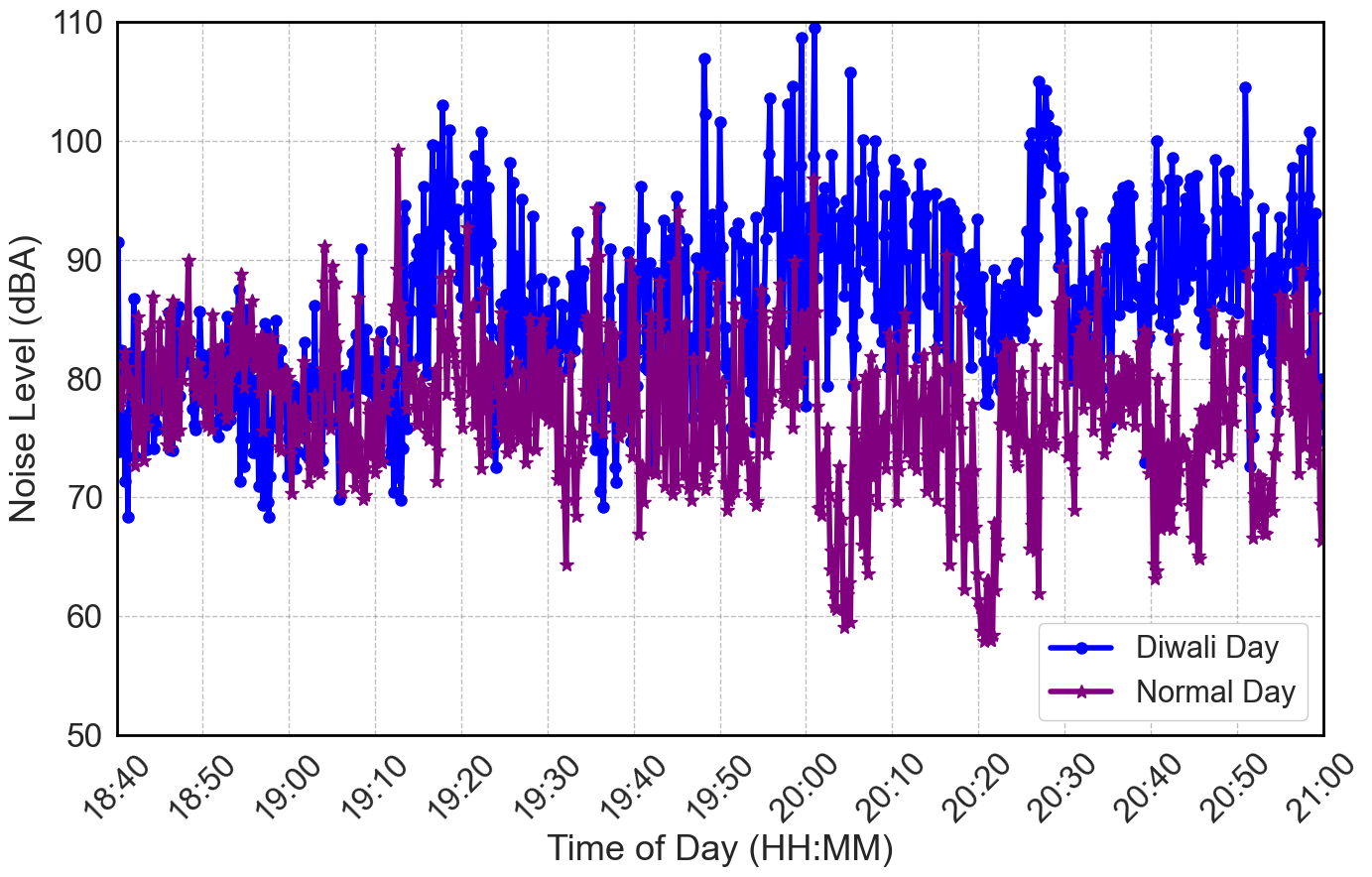}  
        \caption{Time series plot of Diwali vs typical day}
        \label{fig:diwali_timeser}
    \end{subfigure}
    \hfill
    \begin{subfigure}{0.49\textwidth}
        \centering        \includegraphics[width=\textwidth]{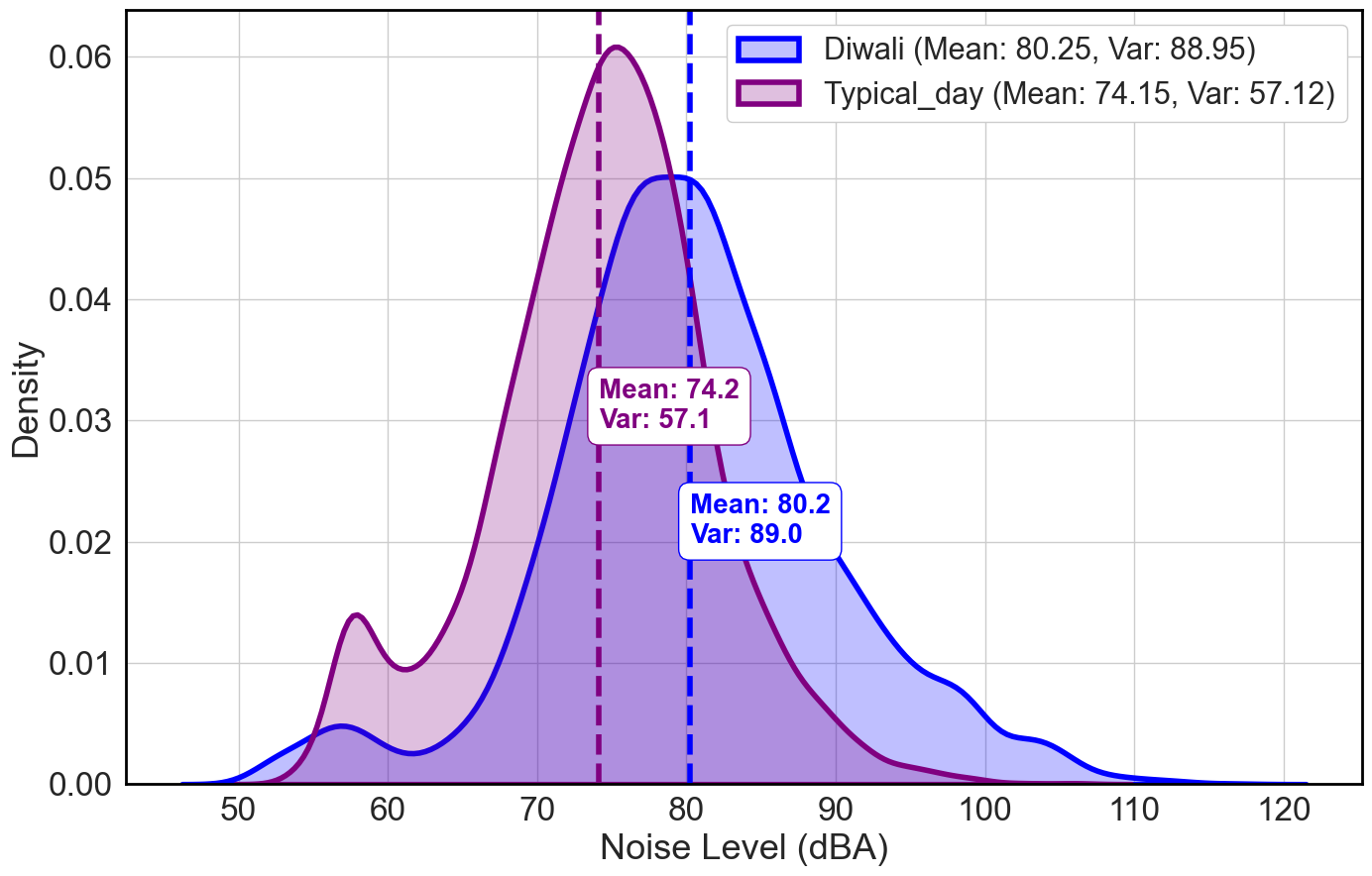}  
        \caption{Noise distribution of Diwali vs typical day}
        \label{fig:diwali_distribution}
    \end{subfigure}
    \hfill
    \caption{Time series and distribution plots comparing noise levels on Diwali and a typical day (Best viewed in color).}
    \label{fig:diwali_noise_dist}
\end{figure*}

\subsection{MC-III}
This measurement campaign was conducted during the Diwali festival to assess the intensity of noise pollution compared to a typical day.  Fig. \ref{fig:diwali_timeser} presents the time series plot, highlighting the difference in noise levels between Diwali and a typical day. An apparent variation emerges from 7 PM, coinciding with the start of bursting crackers during Diwali. The noise levels remain significantly elevated until around 9 PM, primarily due to firecracker bursts. Fig. \ref{fig:diwali_distribution} shows the statistical distribution of the noise data shown in \ref{fig:diwali_timeser}. 
  It is evident that both the mean and variance of noise levels during Diwali celebrations (80.2 and 88.9, respectively) are significantly higher than those of a typical day 74.1 and 57.1). The maximum recorded noise levels reached 110 dBA. The increase in noise levels during Diwali compared to a normal day reflects the intermittent yet intense nature of firecracker noise. Even a few minutes of continuous exposure to noise levels of 110 dBA can pose serious health risks, including temporary or permanent hearing damage, stress, and sleep disturbances for sensitive groups such as children, the elderly, and other vulnerable individuals.  

\begin{figure*}[tbh] 
    \centering
    \setlength{\fboxrule}{0.01pt} 
    \setlength{\fboxsep}{0pt}  

    \begin{subfigure}{0.48\textwidth}
        \centering        \fbox{\includegraphics[width=\linewidth]{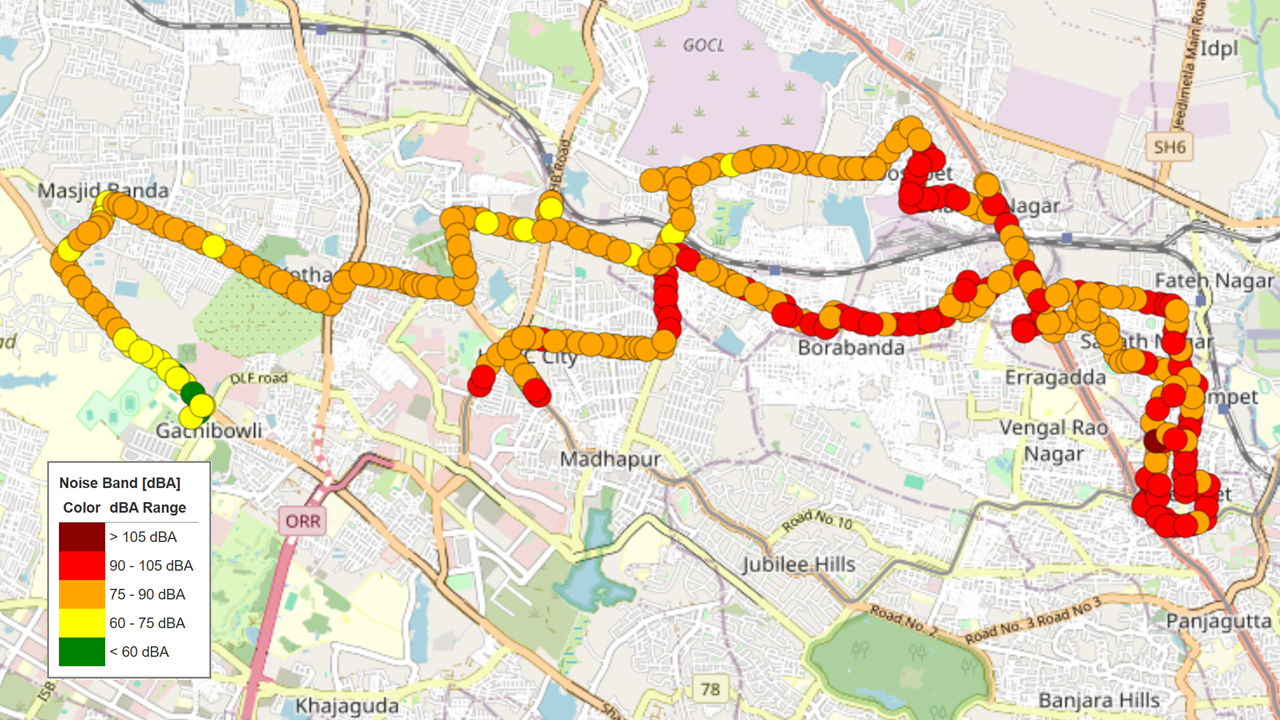}}    
        \caption{Diwali}
    \end{subfigure}
    \hfill
    \begin{subfigure}{0.48\textwidth}
        \centering        \fbox{\includegraphics[width=\linewidth]{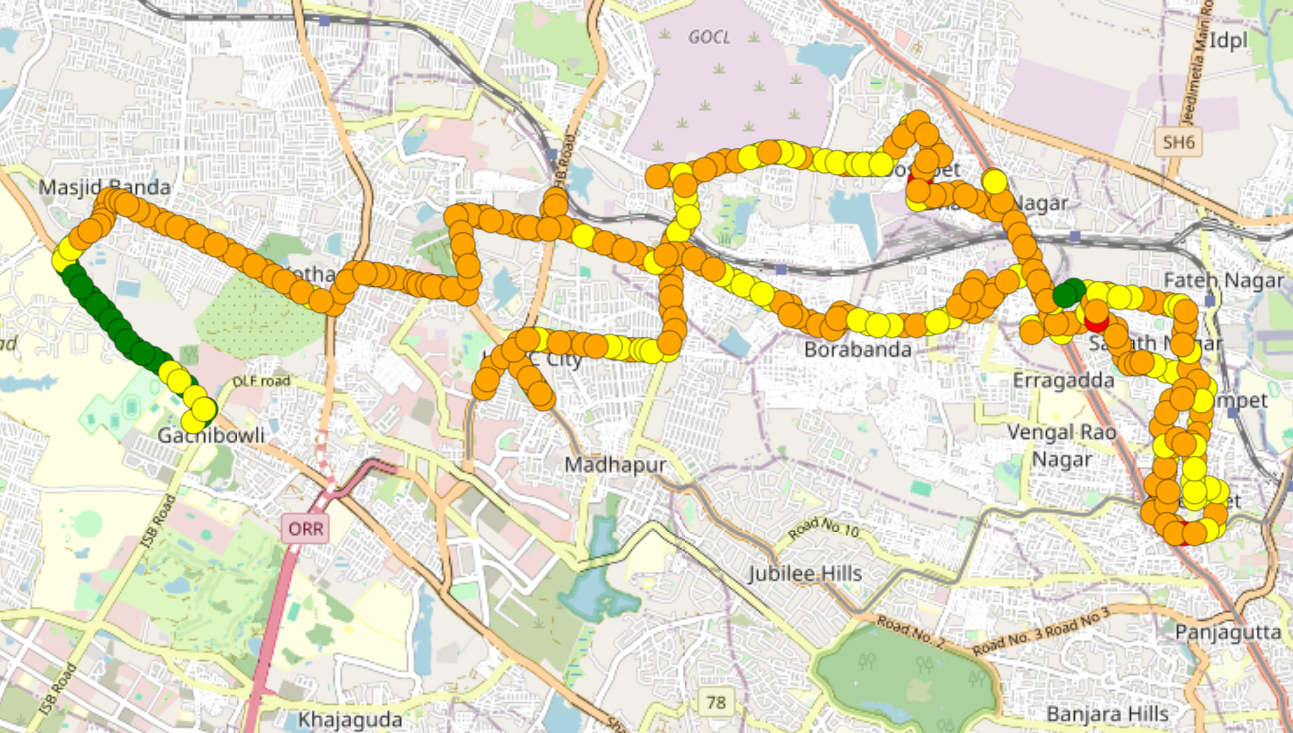}} 
        \caption{Typical day}
    \end{subfigure}
    \caption{Spatial variation of noise levels on Diwali and a typical day (Best viewed in color).}
    \label{fig:diwali_noise_dist_spat}
\end{figure*}

\begin{figure}[bth]
    \centering    \includegraphics[width=0.98\linewidth]{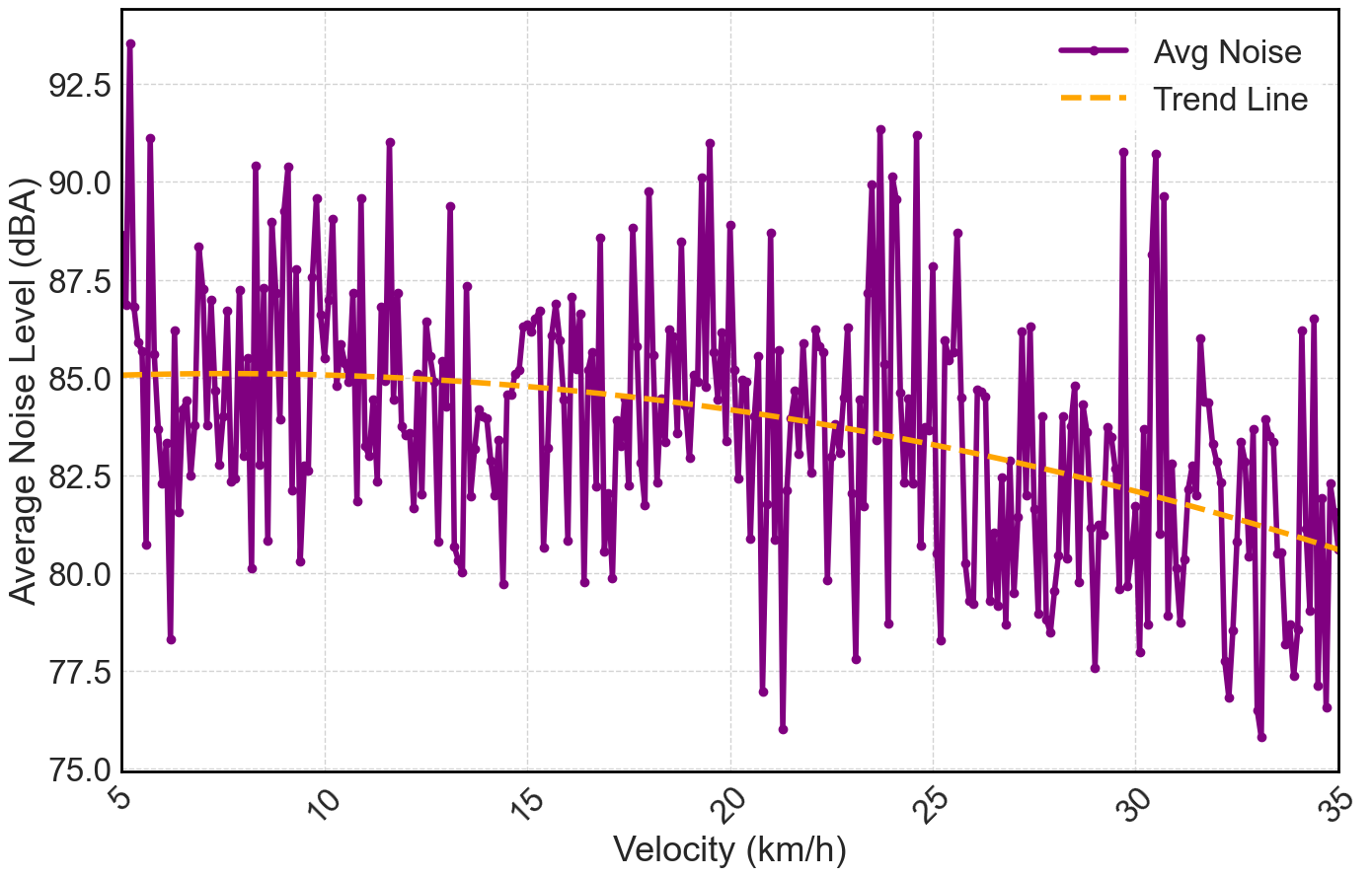}
    \caption{Average noise values as a function of vehicle velocity on Route 3 of MC-II (Best viewed in color).}
    \label{fig:vel}
\end{figure}
Fig. \ref{fig:diwali_noise_dist_spat} provides a spatial noise map for Diwali day and a typical day, depicting the distribution of noise levels in the selected region. The proposed IoT-based solution can help determine hot spots. Higher noise intensity in specific areas indicates concentrated firecracker activity, highlighting the need for potential noise control measures in densely populated residential zones.

\subsection{Effect of traffic on noise levels}
Fig. \ref{fig:vel} illustrates the variation in noise levels of MC-I averaged with respect to vehicle velocity, derived from GPS coordinates and time stamps. It can be seen from the trend line that there is an inverse relationship between the noise values and velocity. Note that lower velocities on the streets during the peak hours are associated with higher traffic congestion. This demonstrates the effect of traffic congestion on the noise values on the streets. This relationship was utilized during the calibration process, resulting in a slight performance improvement, as shown in Table \ref{tab:compar4}. In MLR, velocity is included as a second independent variable, which enhances the calibration metrics compared to the SLR. Similarly, in RFR, adding velocity as an additional feature improved the performance metrics. For example, the R\textsuperscript{2} increased to 0.960 from 0.937, the MAE decreased to 0.693 from 0.85, and the MSE reduced to 0.872 from 1.09.

\section{Discussion}

IoT devices can be effectively used for mobile sensing to monitor noise levels and identify hotspots in highly polluted urban areas. Equipped with sensing components, these devices offer a cost-effective, scalable solution with seamless cloud connectivity, making them suitable for large-scale deployments. 

The use of LCS poses certain challenges, such as the need for calibration, stability, and frequent maintenance. Dynamic testing conditions further emphasize that different sensing nodes may require distinct calibration coefficients for optimal performance. Regarding calibration approaches, there is a trade-off between different performance parameters. For example, SLR and random RFR offer different advantages. SLR-based calibration is computationally more feasible and suitable for controlled environments with stable conditions, such as laboratory chambers or fixed-location data collection, where data variability is less. In contrast, RFR calibration is more effective for uncontrolled environments, particularly in mobile sensing applications, due to its ability to handle variability in the data. While linear regression assumes structured datasets with normally distributed variables, RFR utilizes bin partitioning to accommodate continuous and dynamic data. This capability makes RFR particularly effective for calibrating mobile sensing datasets characterized by high variability in time series measurements. IoT devices calibrated using RFR have shown potential for accurate noise level measurements.

Observations from the data collected across all MCs indicate that most of the surveyed routes are regularly exposed to noise levels exceeding 75 dBA, which is higher than the recommended standards set by the CPCB, as shown in Table~\ref {tab:noise_standards}. While the observations are based on shorter time averages, CPCB standards provide equivalent continuous sound level (Leq) values for day and nighttime periods over longer durations. Nevertheless, the consistently high noise levels observed, exceeding 75 dBA even over short periods, reflect serious concerns regarding urban noise pollution. The elevated short-term noise levels suggest that noise exposure could exceed recommended thresholds when averaged over a full day or night.

To mitigate noise pollution impacts, implementing noise reduction measures, such as creating dense vegetation buffers between roads and residential areas, is recommended. Increasing public awareness among vulnerable commuting groups about the health effects of noise pollution and promoting simple protective measures, such as using earplugs and reducing unnecessary honking, are also important. Expanding the deployment of IoT devices and increasing the frequency of streetcar monitoring runs along identical routes across multiple days would enhance data reliability and provide a more comprehensive understanding of noise patterns.


Despite these findings, the study was limited to a few focused data collection campaigns. Further research is needed to comprehensively understand noise pollution patterns, particularly in heavily affected areas near schools and hospitals. These areas pose heightened risks to vulnerable groups such as children and older adults.

\begin{table}
    \centering
    \small
    \caption{Performance comparison of mobile ML models after including velocity feature}
    \resizebox{0.45\textwidth}{!}{ 
    \begin{tabular}{|l|c|c|c|c|}
        \hline
        \textbf{Method} & \textbf{R$^2$} & \textbf{MAE} & \textbf{RMSE} \\ 
        \hline
        Raw data from node & 0.160 & 23.94 & 24.87 \\ 
        \hline
        Data after pre-processing & 0.690 & 23.86 & 24.07 \\ 
        \hline
        SLR & 0.739 & 1.68 & 2.23 \\ 
        \hline
        MLR (including velocity feature)& \textbf{0.768} & \textbf{1.59} & \textbf{2.10} \\ 
        \hline
        RFR & 0.937 & 0.85 & 1.09 \\ 
        \hline
        RFR (including velocity feature) & \textbf{0.960} & \textbf{0.693} & \textbf{0.872} \\ 
        \hline
        
    \end{tabular}
    }
    \label{tab:compar4}
\end{table}

\begin{table}
    \centering
    \small
    \caption{Ambient air quality standards in respect of noise \cite{noise_rules_2000}.}
    \resizebox{0.49\textwidth}{!}{ 
    \begin{tabular}{|c|l|c|c|}
        \hline
        \textbf{Category of area/zone} & \textbf{Day time (dBA Leq)} & \textbf{Night time (dBA Leq)} \\
        \hline
        Industrial area & 75 & 70 \\
        \hline
         Commercial area & 65 & 55 \\
        \hline
         Residential area & 55 & 45 \\
        \hline
         Silence Zone & 50 & 40 \\
        \hline
    \end{tabular}
    }
    \label{tab:noise_standards}
\end{table}

\section{Conclusion}


This research proposes an IoT-based noise monitoring solution using low-cost sensors stationed on a moving vehicle. 
It is shown that the LCS needs to be calibrated before they are deployed. While laboratory calibration yields promising results, its limited generalizability underscores the necessity for in-field calibration using mobile data from real-world conditions. Among the evaluated ML models for calibration, RFR, particularly when augmented with vehicular velocity, proved most effective for capturing dynamic urban noise variations with performance metrics of \(\textrm{R}^2=0.960\), MAE = 0.693, and RMSE = 0.872.


The analysis on the three MCs conducted across Hyderabad across 27 days, capturing 436,420 data points, revealed significant noise variations linked to different routes, times, and festival events, with average noise levels exceeding 75 dBA and several instances exceeding 90 dBA. These findings emphasize the widespread nature of urban noise pollution and its potential health impacts on residents. The insights from this work underscore the importance of noise pollution monitoring using a mobile IoT node. Mobile monitoring provides higher temporal and spatial resolution, offering valuable data to inform effective noise mitigation strategies in rapidly growing urban areas.

\bibliographystyle{IEEEtran}  
\bibliography{References}  

\end{document}